\documentclass[twocolumn,letterpaper,aps,prc,longbibliography,superscriptaddress,nofootinbib,floatfix]{revtex4-2}
\usepackage{graphicx}	% Include figure files
\usepackage{gensymb}
\usepackage{tabularx}
\usepackage{multirow}
\usepackage{amsmath}

\usepackage{xspace}	% Include xspace

\newcommand{\pt}{\mbox{$p_T$}\xspace}

\newcommand{\Npart}{\mbox{$N_{\rm part}$}\xspace}

\newcommand{\sqsn}{\mbox{$\sqrt{s_{_{NN}}}$}\xspace}
\newcommand{\pp}{\mbox{$p$$+$$p$}\xspace}

\newcommand{\auau}{\mbox{Au$+$Au}\xspace}
\newcommand{\cucu}{\mbox{Cu$+$Cu}\xspace}
\newcommand{\cuau}{\mbox{Cu$+$Au}\xspace}
\newcommand{\uu}{\mbox{U$+$U}\xspace}

\newcommand{\pio}{\mbox{$\pi^0$}\xspace}

\newcommand{\vtwo}{\mbox{$v_2$}\xspace}
\newcommand{\ecc}{\mbox{$\varepsilon_2$}\xspace}
\newcommand{\npart}{\mbox{$N_{\rm part}^{1/3}$}\xspace}
\newcommand{\hydroscaling}{\mbox{$\varepsilon_2 N_{\rm part}^{1/3}$}\xspace}

\newcommand{\deltaphi}{\mbox{$\Delta\varphi$}\xspace}
\newcommand{\rabphi}{\mbox{$R_{AB}(\deltaphi,\pt)$}\xspace}
\newcommand{\slossphi}{\mbox{$S_{\rm loss}(\deltaphi,\pt)$}\xspace}
\newcommand{\psirp}{\mbox{$\Psi^{\rm RP}$}\xspace}
\newcommand{\psiepn}{\mbox{$\Psi_{n}^{\rm EP}$}\xspace}
\newcommand{\psieptwo}{\mbox{$\Psi_{2}^{\rm EP}$}\xspace}
\newcommand{\qn}{\mbox{$Q_{n}$}\xspace}
\newcommand{\qnx}{\mbox{$Q_{n,x}$}\xspace}
\newcommand{\qny}{\mbox{$Q_{n,y}$}\xspace}

\newcommand{\rabincl}{\mbox{$R_{\rm AB}(\pt)$}\xspace}
\newcommand{\slossincl}{\mbox{$S_{\rm loss}(\pt)$}\xspace}

\begin{document}

\title{ Elliptic flow of $\pi^0$ mesons in Cu$+$Au collisions at 
$\sqrt{s_{_{NN}}}=200$~GeV and U$+$U at $\sqrt{s_{_{NN}}}=193$~GeV}

%\author{PHENIX Collaboration}

\newcommand{\abilene}{Abilene Christian University, Abilene, Texas 79699, USA}
\newcommand{\augie}{Department of Physics, Augustana University, Sioux Falls, South Dakota 57197, USA}
\newcommand{\banaras}{Department of Physics, Banaras Hindu University, Varanasi 221005, India}
\newcommand{\barc}{Bhabha Atomic Research Centre, Bombay 400 085, India}
\newcommand{\baruch}{Baruch College, City University of New York, New York, New York, 10010 USA}
\newcommand{\bnlcoll}{Collider-Accelerator Department, Brookhaven National Laboratory, Upton, New York 11973-5000, USA}
\newcommand{\bnlphys}{Physics Department, Brookhaven National Laboratory, Upton, New York 11973-5000, USA}
\newcommand{\caucr}{University of California-Riverside, Riverside, California 92521, USA}
\newcommand{\charlesczech}{Charles University, Faculty of Mathematics and Physics, 180 00 Troja, Prague, Czech Republic}
\newcommand{\ciae}{Science and Technology on Nuclear Data Laboratory, China Institute of Atomic Energy, Beijing 102413, People's Republic of China}
\newcommand{\cns}{Center for Nuclear Study, Graduate School of Science, University of Tokyo, 7-3-1 Hongo, Bunkyo, Tokyo 113-0033, Japan}
\newcommand{\colorado}{University of Colorado, Boulder, Colorado 80309, USA}
\newcommand{\columbia}{Columbia University, New York, New York 10027 and Nevis Laboratories, Irvington, New York 10533, USA}
\newcommand{\czechtech}{Czech Technical University, Zikova 4, 166 36 Prague 6, Czech Republic}
\newcommand{\debrecen}{Debrecen University, H-4010 Debrecen, Egyetem t{\'e}r 1, Hungary}
\newcommand{\elte}{ELTE, E{\"o}tv{\"o}s Lor{\'a}nd University, H-1117 Budapest, P{\'a}zm{\'a}ny P.~s.~1/A, Hungary}
\newcommand{\eszterhazy}{Eszterh\'azy K\'aroly University, K\'aroly R\'obert Campus, H-3200 Gy\"ongy\"os, M\'atrai \'ut 36, Hungary}
\newcommand{\ewha}{Ewha Womans University, Seoul 120-750, Korea}
\newcommand{\fsu}{Florida State University, Tallahassee, Florida 32306, USA}
\newcommand{\gsu}{Georgia State University, Atlanta, Georgia 30303, USA}
\newcommand{\hanyang}{Hanyang University, Seoul 133-792, Korea}
\newcommand{\hiroshima}{Hiroshima University, Kagamiyama, Higashi-Hiroshima 739-8526, Japan}
\newcommand{\hunrenatomki}{HUN-REN ATOMKI, H-4026 Debrecen, Bem t{\'e}r 18/c, Hungary}
\newcommand{\ihepprot}{IHEP Protvino, State Research Center of Russian Federation, Institute for High Energy Physics, Protvino, 142281, Russia}
\newcommand{\illuiuc}{University of Illinois at Urbana-Champaign, Urbana, Illinois 61801, USA}
\newcommand{\inrras}{Institute for Nuclear Research of the Russian Academy of Sciences, prospekt 60-letiya Oktyabrya 7a, Moscow 117312, Russia}
\newcommand{\instpasczech}{Institute of Physics, Academy of Sciences of the Czech Republic, Na Slovance 2, 182 21 Prague 8, Czech Republic}
\newcommand{\isu}{Iowa State University, Ames, Iowa 50011, USA}
\newcommand{\jaea}{Advanced Science Research Center, Japan Atomic Energy Agency, 2-4 Shirakata Shirane, Tokai-mura, Naka-gun, Ibaraki-ken 319-1195, Japan}
\newcommand{\jeonbuk}{Jeonbuk National University, Jeonju, 54896, Korea}
\newcommand{\jyvaskyla}{Helsinki Institute of Physics and University of Jyv{\"a}skyl{\"a}, P.O.Box 35, FI-40014 Jyv{\"a}skyl{\"a}, Finland}
\newcommand{\kek}{KEK, High Energy Accelerator Research Organization, Tsukuba, Ibaraki 305-0801, Japan}
\newcommand{\korea}{Korea University, Seoul 02841, Korea}
\newcommand{\kurchatov}{National Research Center ``Kurchatov Institute", Moscow, 123098 Russia}
\newcommand{\kyoto}{Kyoto University, Kyoto 606-8502, Japan}
\newcommand{\labllr}{Laboratoire Leprince-Ringuet, Ecole Polytechnique, CNRS-IN2P3, Route de Saclay, F-91128, Palaiseau, France}
\newcommand{\lahorelums}{Physics Department, Lahore University of Management Sciences, Lahore 54792, Pakistan}
\newcommand{\lawllnl}{Lawrence Livermore National Laboratory, Livermore, California 94550, USA}
\newcommand{\losalamos}{Los Alamos National Laboratory, Los Alamos, New Mexico 87545, USA}
\newcommand{\lund}{Department of Physics, Lund University, Box 118, SE-221 00 Lund, Sweden}
\newcommand{\maryland}{University of Maryland, College Park, Maryland 20742, USA}
\newcommand{\mass}{Department of Physics, University of Massachusetts, Amherst, Massachusetts 01003-9337, USA}
\newcommand{\mate}{MATE, Institute of Technology, Laboratory of Femtoscopy, K\'aroly R\'obert Campus, H-3200 Gy\"ongy\"os, M\'atrai \'ut 36, Hungary}
\newcommand{\michigan}{Department of Physics, University of Michigan, Ann Arbor, Michigan 48109-1040, USA}
\newcommand{\miss}{Mississippi State University, Mississippi State, Mississippi 39762, USA}
\newcommand{\muhlenberg}{Muhlenberg College, Allentown, Pennsylvania 18104-5586, USA}
\newcommand{\myongji}{Myongji University, Yongin, Kyonggido 449-728, Korea}
\newcommand{\nagasaki}{Nagasaki Institute of Applied Science, Nagasaki-shi, Nagasaki 851-0193, Japan}
\newcommand{\nara}{Nara Women's University, Kita-uoya Nishi-machi Nara 630-8506, Japan}
\newcommand{\natmephi}{National Research Nuclear University, MEPhI, Moscow Engineering Physics Institute, Moscow, 115409, Russia}
\newcommand{\newmex}{University of New Mexico, Albuquerque, New Mexico 87131, USA}
\newcommand{\nmsu}{New Mexico State University, Las Cruces, New Mexico 88003, USA}
\newcommand{\northcg}{Physics and Astronomy Department, University of North Carolina at Greensboro, Greensboro, North Carolina 27412, USA}
\newcommand{\ohio}{Department of Physics and Astronomy, Ohio University, Athens, Ohio 45701, USA}
\newcommand{\ornl}{Oak Ridge National Laboratory, Oak Ridge, Tennessee 37831, USA}
\newcommand{\orsay}{IPN-Orsay, Univ.~Paris-Sud, CNRS/IN2P3, Universit\'e Paris-Saclay, BP1, F-91406, Orsay, France}
\newcommand{\pnpi}{PNPI, Petersburg Nuclear Physics Institute, Gatchina, Leningrad region, 188300, Russia}
\newcommand{\riken}{RIKEN Nishina Center for Accelerator-Based Science, Wako, Saitama 351-0198, Japan}
\newcommand{\rikjrbrc}{RIKEN BNL Research Center, Brookhaven National Laboratory, Upton, New York 11973-5000, USA}
\newcommand{\rikkyo}{Physics Department, Rikkyo University, 3-34-1 Nishi-Ikebukuro, Toshima, Tokyo 171-8501, Japan}
\newcommand{\saispbstu}{Saint Petersburg State Polytechnic University, St.~Petersburg, 195251 Russia}
\newcommand{\seoulnat}{Department of Physics and Astronomy, Seoul National University, Seoul 151-742, Korea}
\newcommand{\stonybrkc}{Chemistry Department, Stony Brook University, SUNY, Stony Brook, New York 11794-3400, USA}
\newcommand{\stonycrkp}{Department of Physics and Astronomy, Stony Brook University, SUNY, Stony Brook, New York 11794-3800, USA}
\newcommand{\sungskku}{Sungkyunkwan University, Suwon, 440-746, Korea}
\newcommand{\tenn}{University of Tennessee, Knoxville, Tennessee 37996, USA}
\newcommand{\titech}{Department of Physics, Tokyo Institute of Technology, Oh-okayama, Meguro, Tokyo 152-8551, Japan}
\newcommand{\tsukuba}{Tomonaga Center for the History of the Universe, University of Tsukuba, Tsukuba, Ibaraki 305, Japan}
\newcommand{\usmma}{United States Merchant Marine Academy, Kings Point, New York 11024, USA}
\newcommand{\vandy}{Vanderbilt University, Nashville, Tennessee 37235, USA}
\newcommand{\weizmann}{Weizmann Institute, Rehovot 76100, Israel}
\newcommand{\wigner}{Institute for Particle and Nuclear Physics, HUN-REN Wigner Research Centre for Physics, (HUN-REN Wigner RCP, RMI), H-1525 Budapest 114, POBox 49, Budapest, Hungary}
\newcommand{\yonsei}{Yonsei University, IPAP, Seoul 120-749, Korea}
\newcommand{\zagreb}{Department of Physics, Faculty of Science, University of Zagreb, Bijeni\v{c}ka c.~32 HR-10002 Zagreb, Croatia}
\newcommand{\zambia}{Department of Physics, School of Natural Sciences, University of Zambia, Great East Road Campus, Box 32379, Lusaka, Zambia}
\affiliation{\abilene}
\affiliation{\augie}
\affiliation{\banaras}
\affiliation{\barc}
\affiliation{\baruch}
\affiliation{\bnlcoll}
\affiliation{\bnlphys}
\affiliation{\caucr}
\affiliation{\charlesczech}
\affiliation{\ciae}
\affiliation{\cns}
\affiliation{\colorado}
\affiliation{\columbia}
\affiliation{\czechtech}
\affiliation{\debrecen}
\affiliation{\elte}
\affiliation{\ewha}
\affiliation{\fsu}
\affiliation{\gsu}
\affiliation{\hanyang}
\affiliation{\hiroshima}
\affiliation{\hunrenatomki}
\affiliation{\ihepprot}
\affiliation{\illuiuc}
\affiliation{\inrras}
\affiliation{\instpasczech}
\affiliation{\isu}
\affiliation{\jaea}
\affiliation{\jeonbuk}
\affiliation{\jyvaskyla}
\affiliation{\kek}
\affiliation{\korea}
\affiliation{\kurchatov}
\affiliation{\kyoto}
\affiliation{\labllr}
\affiliation{\lahorelums}
\affiliation{\lawllnl}
\affiliation{\losalamos}
\affiliation{\lund}
\affiliation{\maryland}
\affiliation{\mass}
\affiliation{\mate}
\affiliation{\michigan}
\affiliation{\miss}
\affiliation{\muhlenberg}
\affiliation{\myongji}
\affiliation{\nagasaki}
\affiliation{\nara}
\affiliation{\natmephi}
\affiliation{\newmex}
\affiliation{\nmsu}
\affiliation{\northcg}
\affiliation{\ohio}
\affiliation{\ornl}
\affiliation{\orsay}
\affiliation{\pnpi}
\affiliation{\riken}
\affiliation{\rikjrbrc}
\affiliation{\rikkyo}
\affiliation{\saispbstu}
\affiliation{\seoulnat}
\affiliation{\stonybrkc}
\affiliation{\stonycrkp}
\affiliation{\sungskku}
\affiliation{\tenn}
\affiliation{\titech}
\affiliation{\tsukuba}
\affiliation{\usmma}
\affiliation{\vandy}
\affiliation{\weizmann}
\affiliation{\wigner}
\affiliation{\yonsei}
\affiliation{\zagreb}
\affiliation{\zambia}
\author{N.J.~Abdulameer} \affiliation{\debrecen} \affiliation{\hunrenatomki}
\author{U.~Acharya} \affiliation{\gsu}
\author{C.~Aidala} \affiliation{\losalamos} \affiliation{\michigan} 
\author{N.N.~Ajitanand} \altaffiliation{Deceased} \affiliation{\stonybrkc} 
\author{Y.~Akiba} \email[PHENIX Spokesperson: ]{akiba@rcf.rhic.bnl.gov} \affiliation{\riken} \affiliation{\rikjrbrc}
\author{R.~Akimoto} \affiliation{\cns} 
\author{J.~Alexander} \affiliation{\stonybrkc} 
\author{D.~Anderson} \affiliation{\isu}
\author{S.~Antsupov} \affiliation{\saispbstu}
\author{K.~Aoki} \affiliation{\kek} \affiliation{\riken} 
% ...
\collaboration{PHENIX Collaboration}  \noaffiliation

\date{\today}

%-----------------------------------------------------------------------------|

\begin{abstract}

%\linenumbers

The second-order azimuthal anisotropy coefficients ($v_2$) of neutral $\pi$ 
mesons ($\pi^0$) have been measured as a function of the transverse 
momentum ($p_T$) and centrality of Cu$+$Au collisions at 
$\sqrt{s_{_{NN}}}=200$~GeV and U$+$U at $\sqrt{s_{_{NN}}}=193$ GeV at the 
Relativistic Heavy Ion Collider. The analysis used experimental data 
collected by the PHENIX experiment at midrapidity $|\eta|<0.35$ over a 
broad $p_T$ range up to $\approx10$~GeV/$c$, and the obtained results 
are compared with previous PHENIX measurements in Au$+$Au collisions at 
$\sqrt{s_{_{NN}}}=200$~GeV. In all three collision systems, 
the $\pi^0$~$v_2$ values follow the scaling with the second-order 
participant eccentricity and the cube root of the number of 
participating nucleons ($\varepsilon_2 N_{\rm part}^{1/3}$) up to 
$\approx4$~GeV/$c$. Furthermore, the behavior of the azimuthal-dependent 
$\pi^0$ nuclear-modification factors and associated fractional 
parton-energy losses are evaluated from measured nonzero $v_2$ values of 
$\pi^0$ at $p_T>5$ GeV/$c$ and found to be approximately the same for 
similar values of $N_{\rm part}^{1/3}$ in these collision systems. These 
findings demonstrate that the mechanism of $\pi^0$ $v_2$ generation 
exhibits a high degree of universality across different initial 
geometries of heavy-ion collisions.

\end{abstract}

\pacs{25.75.Dw}

\maketitle

\section{INTRODUCTION}

The unique state of matter consisting of deconfined partons is called 
the quark-gluon plasma (QGP), and it can be created in 
ultra-relativistic heavy-ion collisions~\cite{QGP1, QGP2}. Determining 
the properties of the QGP remains a central goal of high-energy nuclear 
physics. One of the observables, which allows the characterization of 
collective behavior and the transport properties of the QGP, is the 
azimuthal anisotropy of final-state particle 
distributions~\cite{AzimuthalAnisotropy}. This is usually quantified by 
the coefficients $v_n$ of the Fourier expansion of the azimuthal 
particle distribution in momentum space relative to the reaction plane 
(the plane spanned by the beam axis and the impact parameter vector). 
The second Fourier coefficient, \vtwo, quantifies the elliptic 
anisotropy and is typically the dominant harmonic at 
midrapidity~\cite{CollectivePhenomena}.

One of the ways to investigate the elliptic flow is to measure its 
\pt-dependence, which allows the determination of the dominant mechanism 
of \vtwo development. Recent studies~\cite{Heinz:2013th} show that the 
elliptic-flow values at low transverse momenta ($\pt<$~2--3~GeV/$c$) are 
successfully described by relativistic hydrodynamics, considering QGP as 
a near-perfect fluid with a low ratio of shear viscosity to entropy 
density. However, at higher transverse momenta ($\pt>5$ GeV/$c$), where 
hard processes start to dominate, the observed nonzero \vtwo values 
indicate path-length-dependent parton-energy 
losses~\cite{Gyulassy:2000gk}. This feature makes \vtwo a unique 
quantity that enables the simultaneous study of the collective behavior 
and the transport properties of the QGP. Previous measurements of 
elliptic flow of charged hadrons and $\ensuremath{\phi}$ mesons in 
symmetric and asymmetric large collision systems (\cucu, \auau, \cuau, 
\uu~\cite{PPG124, PPG183, PPGYura}) revealed a scaling of \vtwo up to 
$p_T\approx4$ GeV/$c$ with a second-order participant eccentricity and a 
cube root of the participant nucleon number (\hydroscaling). This 
scaling is consistent with the predictions of the relativistic 
hydrodynamic model of QGP~\cite{Taranenko}, suggesting that the 
collective expansion of QGP is the main driver of elliptic flow. On the 
other hand, the nonzero \vtwo values of \pio observed at $p_T>5$ GeV/$c$ 
in Au$+$Au collisions have been interpreted as reflecting the 
path-length dependence of parton-energy loss~\cite{Gyulassy:2000gk, 
Wang2001, Gyulassy:2003mc}. To study the behavior of the elliptic flow 
at high values of \pt, the \pio nuclear-modification factors (\rabphi) 
and the associated fractional parton-energy losses (\slossphi) were 
measured in \auau collisions as a function of \pt and \pio emission 
angle (\deltaphi) relative to the reaction plane~\cite{QGP2, 
PRC_pi0_AUAU_2007}. The relatively small uncertainties associated with 
\pio production in heavy-ion collisions over a wide range of 
\pt~\cite{ZHARKO, RADZEVICH} enable a detailed study of \vtwo$(\pt)$.

Because the elliptic flow reflects how the initial anisotropy of the 
nuclear-overlap region is converted into the momentum anisotropy of 
final-state particles, the \vtwo values quantify the medium's response 
to the initial spatial anisotropy. Due to the fact that the initial 
geometry changes as a function of centrality, measuring the \vtwo 
values across several centrality bins in different collision systems 
sheds light on the dependence of elliptic flow on the initial collision 
geometry. While \vtwo values of \pio have been extensively studied as a 
function of centrality in symmetric Au$+$Au 
collisions~\cite{PRC_pi0_AUAU_2007,PRC_pi0_AUAU_2009,PRC_pi0_AUAU_2010,PPG129}, 
measurements in asymmetric systems such as \cuau or collisions of deformed 
nuclei such as \uu can be used for additional systematic study of how the 
initial conditions influence the development of \pio \vtwo. The 
nuclear-overlap region in \cuau collisions has additional asymmetry in 
the reaction plane compared to \auau collisions due to the different 
sizes of copper and gold nuclei. In contrast, \uu collisions are 
symmetric in mass number, but comprise prolate deformed $^{238}$U 
nuclei. At the time of collision, the nuclei are randomly oriented with 
respect to each other and, therefore, with respect to the reaction 
plane. For that reason, the shape and size of nuclear-overlap region 
can vary significantly within the same centrality 
bin~\cite{U_orientation}, unlike symmetric collisions of spherical or 
nearly spherical nuclei such as \auau. Consequently, even the most 
central U$+$U collisions can have an azimuthally anisotropic shape, which 
makes this collision system especially interesting from the point of 
view of studying the path-length dependence of jet 
quenching~\cite{U_quenching}. Thus, examining the \hydroscaling scaling 
for \pio \vtwo values and estimating \pio \rabphi and \slossphi in 
asymmetric and deformed collision systems such as \cuau and \uu 
collisions, in comparison with symmetric \auau collisions, provides the 
opportunity to check the universality of the \pio \vtwo generation 
mechanisms across different initial-collision geometries.

Presented here are \vtwo measurements of \pio mesons in \cuau collisions 
at \sqsn~=~200 GeV and \uu at \sqsn~=~193 GeV, using data collected by 
the PHENIX experiment at the Relativistic Heavy Ion Collider in 2012. 
The azimuthal dependences of \pio nuclear-modification factors and the 
associated fractional parton-energy losses are estimated in these 
collision systems using the measured values of \pio \vtwo at $p_T>5$ 
GeV/$c$.

%%%%%%%%%%%%%%%%%%%%%%%%%%%%%%%%%%%%%%%%%%%%%%%%%%%%%%%%%%%%%%%%%%%%%%%%%%

\section{DATA ANALYSIS}

\subsection{Data sets and event categorization}

\begin{figure}[]
\includegraphics[width=1.0\linewidth]{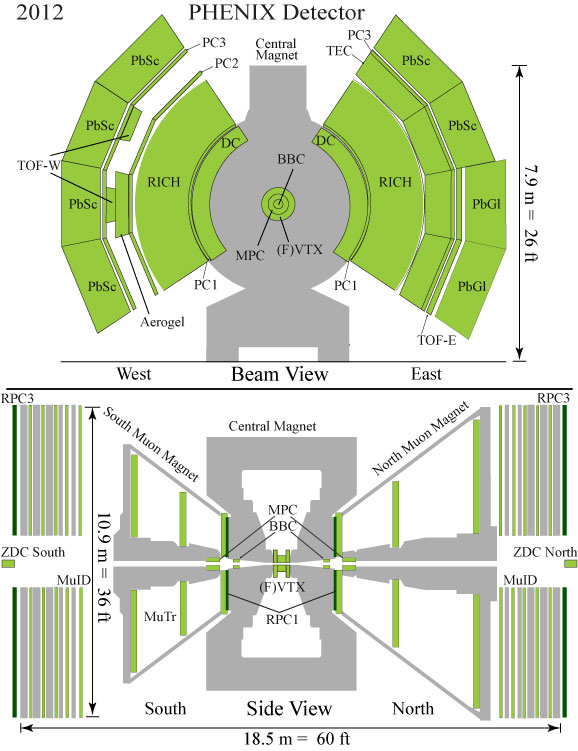}
\caption{PHENIX experimental setup in the 2012 operational period.}
\label{fig:detector}
\end{figure}

The measurements presented in this work are based on \cuau collision 
data at \sqsn~=~200 GeV and \uu collision data at \sqsn~=~193 GeV 
collected by the PHENIX experiment during the 2012 operational period. 
The experimental setup is shown in Fig.~\ref{fig:detector}.

%%%%%%%%%%%%%%%%%%%%%%%%%%%%%%%%%%%%%%%%%%%%%%%%%%%%%%% Fig_1

The categorization of an event is carried out by determining its 
centrality and collision vertex. Both of these are 
determined using the beam-beam counters 
(BBC)~\cite{PHENIX_InnerDetectors}. The BBC comprises two arms (north 
and south), one on either side of the experiment, located at 
$z=\pm144~{\rm cm}$ from the nominal center of the experiment and 
having a pseudorapidity acceptance of $3.0<|\eta|<3.9$. The centrality 
is estimated using the multiplicity of charged particles registered by 
the BBC. The vertex is determined using the time difference between the 
north and the south arms of the BBC.

\subsection{Reaction plane} 

In each collision, a specific plane can be identified. The reaction 
plane is defined by the beam axis and the vector of the impact 
parameter. The orientation of the reaction plane relative to the 
coordinate system of the experiment is characterized by its azimuthal 
angle (\psirp). Because the impact parameter vector is not 
experimentally accessible, the angle \psirp cannot be measured directly. 
Instead, an event-plane angle ($\psiepn$) is calculated for each 
harmonic number $n$ using the flow vector 
\qn~\cite{Methods_for_analyzing}. The vector \qn characterizes the 
$n$-th order anisotropic flow direction in the event, which can be 
calculated as:
%\begin{linenomath} 
\begin{equation}
    \label{eq:qn}
        \qn = \qnx + i\qny = |Q_n|e^{i(n\cdot\psiepn)},
\end{equation}
%\end{linenomath}
where $\qnx = \sum_{i}^{N}{\omega_{i}\cos(n\varphi_{i})}$, $\qny = 
\sum_{i}^{N}{\omega_{i}\sin(n\varphi_{i})}$, $\psiepn$ is the $n$-th 
order event-plane angle, $N$ is the number of particles in the event, 
$\varphi_{i}$ is the azimuthal angle of a particle in the detector 
frame, and $\omega_{i}$ is the weight for these particles (for example, 
\pt~\cite{Methods_for_analyzing}). The angle \psiepn is estimated by 
measuring the azimuthal angles of particles that form the vector \qn and 
can be calculated as
%\begin{linenomath}
\begin{equation}
    \label{eq:psiepn}
    \psiepn=\frac{1}{n}\arctan\left(\frac{\qny}{\qnx}\right).
\end{equation}
%\end{linenomath}

However, nonuniform azimuthal acceptance of the detector leads to 
additional correlations when measuring \qn vectors resulting in 
incorrect estimation of the \psiepn angles.  Used here are the standard 
procedures to correct for nonuniform acceptance, which include 
recentering the \qn vectors~\cite{Methods_for_analyzing} and then 
flattening the \psiepn angles~\cite{Flattening}.

Due to finite sampling, the event-plane angle is not perfectly 
determined, which reduces the strength of the measured angular 
correlations. Measured $v_n$ values must therefore be corrected by the 
event-plane resolution~(Res)~\cite{Methods_for_analyzing}, given by
%\begin{linenomath}
\begin{equation}
    \label{eq:res}
    \text{Res}\{\psiepn\}=\langle\cos[n(\psiepn-\psirp)]\rangle,
\end{equation}
%\end{linenomath}
where the angle brackets denote the average over all events. 
Additionally, it is necessary to have a large pseudorapidity gap between 
the measurements of \pio \vtwo ($|\eta|<0.35$) and the determination of 
the event plane in order to reduce nonflow correlations between these 
measurements, which is especially important for high-\pt particles and 
low multiplicity events~\cite{DirPhotAuAu}.

In the \cuau analysis, the event plane is determined using the forward 
silicon-vertex detector (FVTX)~\cite{FVTX} for \pio with $p_T<5~{\rm 
GeV}/c$ and using the BBC for \pio with $p_T\ge 5~{\rm GeV}/c$. The 
FVTX detector provides accurate charged-particle-tracking measurements 
and, therefore, has better resolution than the BBC. However, the 
pseudorapidity gap between the FVTX and the PHENIX central arms 
(CNT)~\cite{TrackingSystem} where the \pio is measured is smaller than 
that between the BBC and the CNT, making measurements with the FVTX 
potentially more susceptible to nonflow. Nonflow is more prevalent at 
higher $p_T$. Thus, the BBCs are used for the event plane determination 
for higher $p_T$ \pio. In the low-to-intermediate transverse-momentum 
range $\pt<5$ GeV/$c$, where the $v_2$ is primarily driven by the 
elliptic flow from the hydrodynamic expansion of the QGP, the 
difference between the \vtwo measurements using the FVTX and the BBC is 
negligible~\cite{DirPhotAuAu}. In the \uu analysis, the muon-piston 
calorimeter (MPC)~\cite{MPC} is used for all measurements. The MPC is a 
lead-tungstate calorimeter that provides simultaneous measurements of 
charged and neutral particles, has a finer granularity than the BBC, 
and covers almost the same pseudorapidity range (the southern part of 
the MPC covers $-3.7<\eta<-3.1$, the northern part of the MPC covers 
$3.1<\eta<3.9$). The MPC has better event-plane resolution than the BBC 
and provides the same pseudorapidity gap.

The event-plane resolution values for each detector are calculated 
using the three-subevent method~\cite{Methods_for_analyzing}, where 
three sets of correlations between independent \psieptwo measurements 
are used to calculate the resolution. For the \cuau analysis, the BBC, 
FVTX, and CNT are used to determine the resolution. For the \uu 
analysis, the BBC, MPC, and CNT are used to determine the resolution. The 
evaluated resolution Res$\{\Psi_{2}^{\rm EP}\}$ values used in this 
work are presented in Table~\ref{table:res}.

%========================================================== Table_I
\begin{table}[tb]
\caption{Resolution of \psieptwo measured in the BBC, FVTX and MPC at 
10\%-wide centrality bins of \cuau collisions at \sqsn~=~200 GeV and \uu 
collisions at \sqsn~=~193 GeV.}
\label{table:res}
\begin{ruledtabular}
\begin{tabular}{cccc} % Changed to 6 columns
Centrality & \multicolumn{2}{c}{\cuau} & \uu \\
 & Res$\{\Psi_{2}^{\rm BBC}\}$ & Res$\{\Psi_{2}^{\rm FVTX}\}$ 
 & Res$\{\Psi_{2}^{\rm MPC}\}$ \\
\hline
0\%--10\% & 0.194 &  0.350  & 0.439 \\
10\%--20\% & 0.228 &  0.396  & 0.567 \\
20\%--30\% & 0.240 &  0.415  & 0.573 \\
30\%--40\% & 0.221 &  0.393  & 0.502 \\
40\%--50\% & 0.181 &  0.339  & 0.385 \\
50\%--60\% & 0.133 &  0.264  & 0.249 \\
\end{tabular}
\end{ruledtabular}
\end{table}

\subsection{Neutral pions}

Reconstruction of \pio mesons is performed via the two-photon decay 
mode, $\pio\rightarrow{\gamma\gamma}$. Photons are measured by the 
PHENIX CNT electromagnetic calorimeter (EMCal)~\cite{EMCal} covering 
$|\eta|<0.35$ and consisting of 8 sectors: 6 lead-scintillator sectors 
(PbSc) and 2 lead-glass \v{C}erenkov sectors (PbGl).

%%%%%%%%%%%%%%%%%%%%%%%%%%%%%%%%%%%%%%%%%%%%%%%%%%%%%%% Fig_2
\begin{figure}[]
\includegraphics[width=1.0\linewidth]{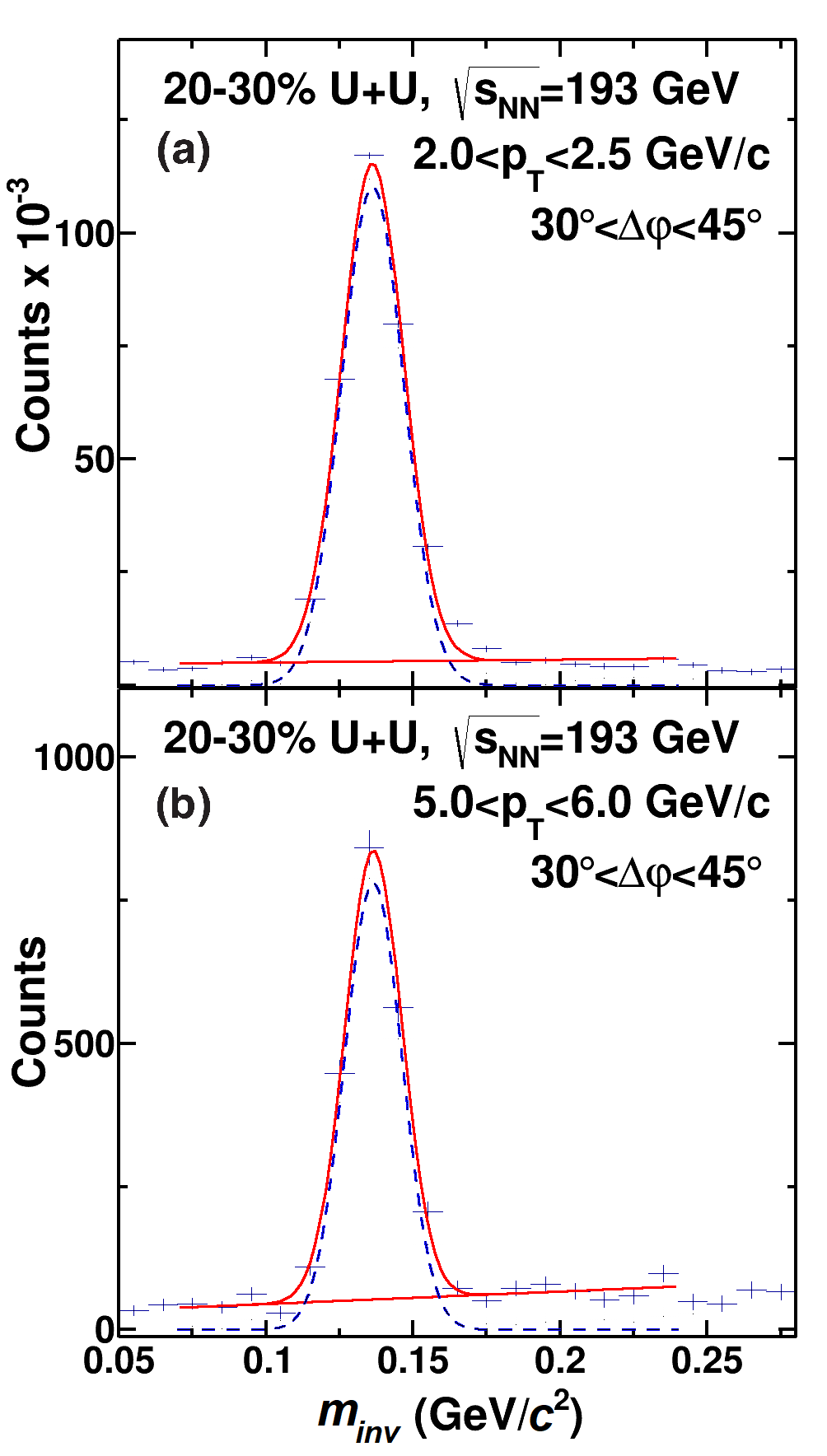}
\caption{Examples of invariant mass distributions after subtracting 
the mixed event backgrounds (20\%--30\% centrality of \uu collisions, 
(a) 2.0~GeV/$c$ $<\pt<2.5$~GeV/$c$ and (b) 5.0~GeV/$c$ $<\pt<6.0$~GeV/$c$ 
with $30\degree<\Delta\varphi<45\degree$). The foreground is 
represented by [blue] histogram-cell crosses, the solid [red] curves are 
Gaussian fits of the signal and the second-order polynomial fits of the 
remaining background, the dashed [blue] curves are Gaussian fits of the pure 
signal (after subtracting the remaining background).}

\label{fig:rawyield}
\end{figure}

The cuts used for each photon and $\gamma\gamma$ pair are the same as 
presented in Ref.~\cite{ZHARKO,RADZEVICH}. Photon candidates are 
required to satisfy the following criteria: $\gamma$ candidates should 
pass the shower-shape cut~\cite{EMCal}, and their energy ($E_{\gamma}$) 
should be more than $0.4$ GeV. The shower-shape cut and energy cut 
reduce the contribution from nonphotonic particles registered in the 
EMCal. Photon pairs are required to be in the same sector of the EMCal 
and the distance between them is required to be more than 
$8$~cm~\cite{PRC_pi0_AUAU_2007}. Photon pairs are also required to pass 
an energy-asymmetry cut $\alpha=|E_{\gamma1}-E_{\gamma2}
|/(E_{\gamma1}+E_{\gamma2})<0.8$, which improves the signal to 
background ratio~\cite{ZHARKO}.

Pairs of photons passing all cuts are combined to construct the 
invariant mass distribution ($m_{\rm inv}$) in bins of \pt, centrality, 
and their azimuthal angle relative to the event plane 
($\Delta\varphi=\varphi-\psieptwo$). The mass spectrum comprises a 
smeared peak around the rest mass of the \pio meson, representing the 
signal part, and the background formed by uncorrelated photon pairs is 
the remaining part. The smearing of the signal peak is due to the finite 
energy and position resolution of the EMCal. To accurately calculate the 
\pio yields, the background part should be subtracted from the observed 
invariant mass spectrum using the mixed-event 
technique~\cite{PRC_pi0_AUAU_2007}. The residual background contribution 
is approximated by a second-order polynomial function and subsequently 
removed from the invariant-mass spectrum. An example of the $m_{\rm inv}$ 
distribution after subtracting the remaining background is shown in 
Fig.~\ref{fig:rawyield}. The \pio yields are calculated by integrating 
the resulting $m_{\rm inv}$ within a range of $\pm2\sigma$ around the 
mean.

\subsection{Elliptic flow}

\begin{figure}[]
\includegraphics[width=1.0\linewidth]{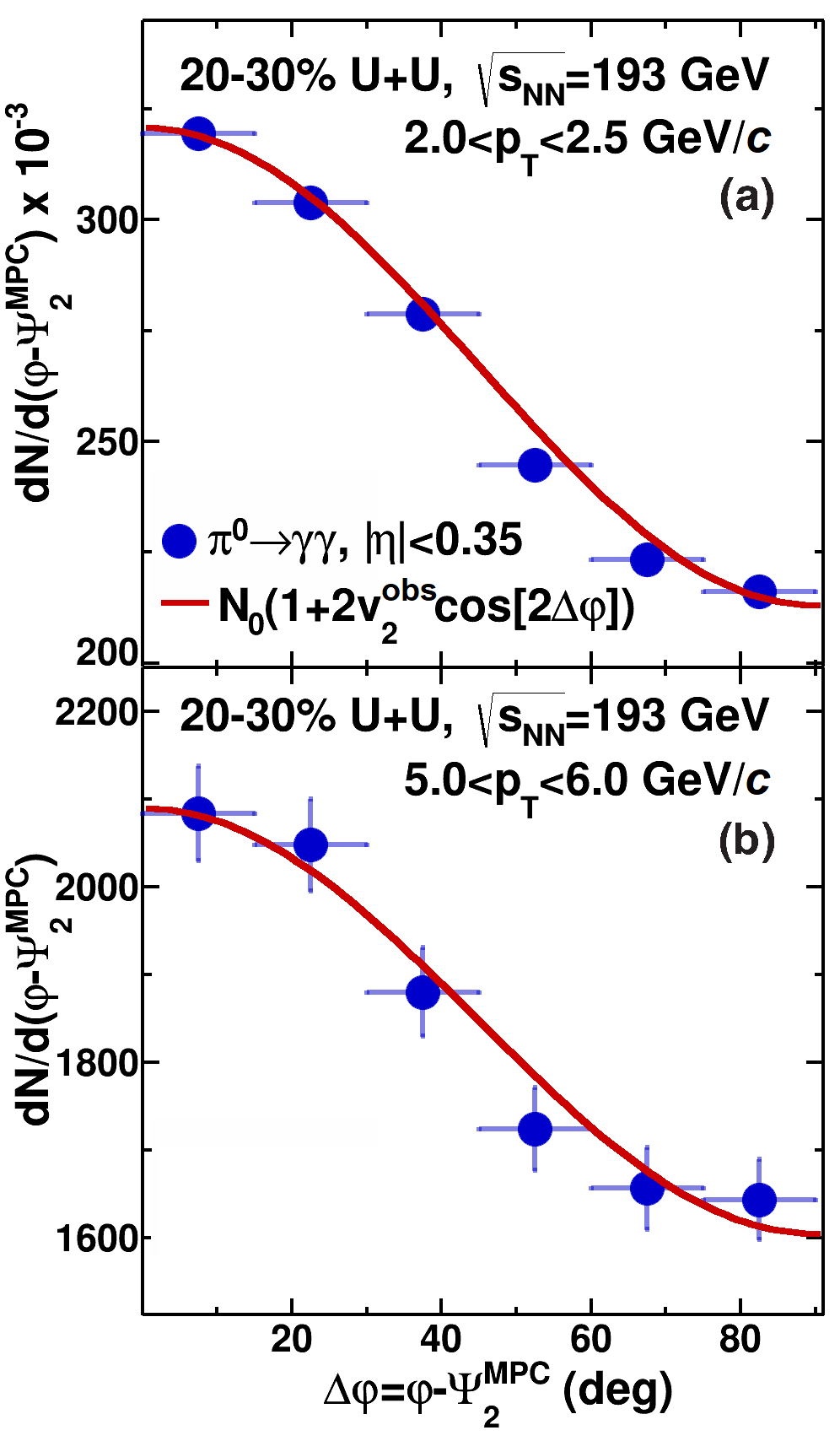}
\caption{Examples of the $dN/d\Delta\varphi$ distributions (20\%--30\% 
centrality of \uu collisions for (a) 2.0~GeV/$c$ $<\pt<2.5$~GeV/$c$ and 
(b) 5.0~GeV/$c$ $<\pt<6.0$~GeV/$c$). The solid [red] line is the function from 
Eq.~\protect\ref{eq:dndphi}.}
\label{fig:dndphi}
\end{figure}

The observed elliptic flow ($v_{2}^{\rm obs}$) of \pio mesons is calculated 
as the second Fourier coefficient of the \pio azimuthal distribution 
relative to the event plane~\cite{Methods_for_analyzing}:
%\begin{linenomath}
\begin{equation}
    \label{eq:vtwo}
    v_{2}^{\rm obs} = \langle\cos[2(\varphi-\psieptwo)]\rangle,
\end{equation}
%\end{linenomath}
where the angle brackets denote an average over all particles and 
events. The \pio yields were measured in six azimuthal angle intervals 
relative to the event plane $(dN/d\Delta\varphi)$ in the range 
$0<\Delta\varphi<\pi/2$. Assuming that the dominant source of \pio 
yields modulation as a function of $\Delta\varphi$ is the elliptic 
flow~\cite{PRC_pi0_AUAU_2009}, the resulting distribution 
$dN/d\Delta\varphi$ can be approximated by
%\begin{linenomath}
\begin{equation}
    \label{eq:dndphi}
    dN/d\Delta\varphi = N_{0}(1+2v_{2}^{\rm obs}\cos[2\Delta\varphi]),
\end{equation}
%\end{linenomath}
where $N_{0}$ is the normalization constant that quantifies the \pio 
yields per each $\Delta\varphi$ interval in the absence of elliptic 
flow. An example of the $dN/d\Delta\varphi$ distribution is shown in 
Fig.~\ref{fig:dndphi}. The observed $v_2$ is corrected by the 
event-plane resolution to yield
%\begin{linenomath}
\begin{equation}
    \label{eq:v2corr}
    v_{2} = v_{2}^{\rm obs}/{\rm Res}\{\psieptwo\}.
\end{equation}
%\end{linenomath}

%%%%%%%%%%%%%%%%%%%%%%%%%%%%%%%%%%%%%%%%%%%%%%%%%%%%%%% Fig_3

%%%%%%%%%%%%%%%%%%%%%%%%%%%%%%%%%%%%%%%%%%%%%%%%%%%%%%% Fig_4
\begin{figure*}[]
\includegraphics[width=0.99\linewidth]{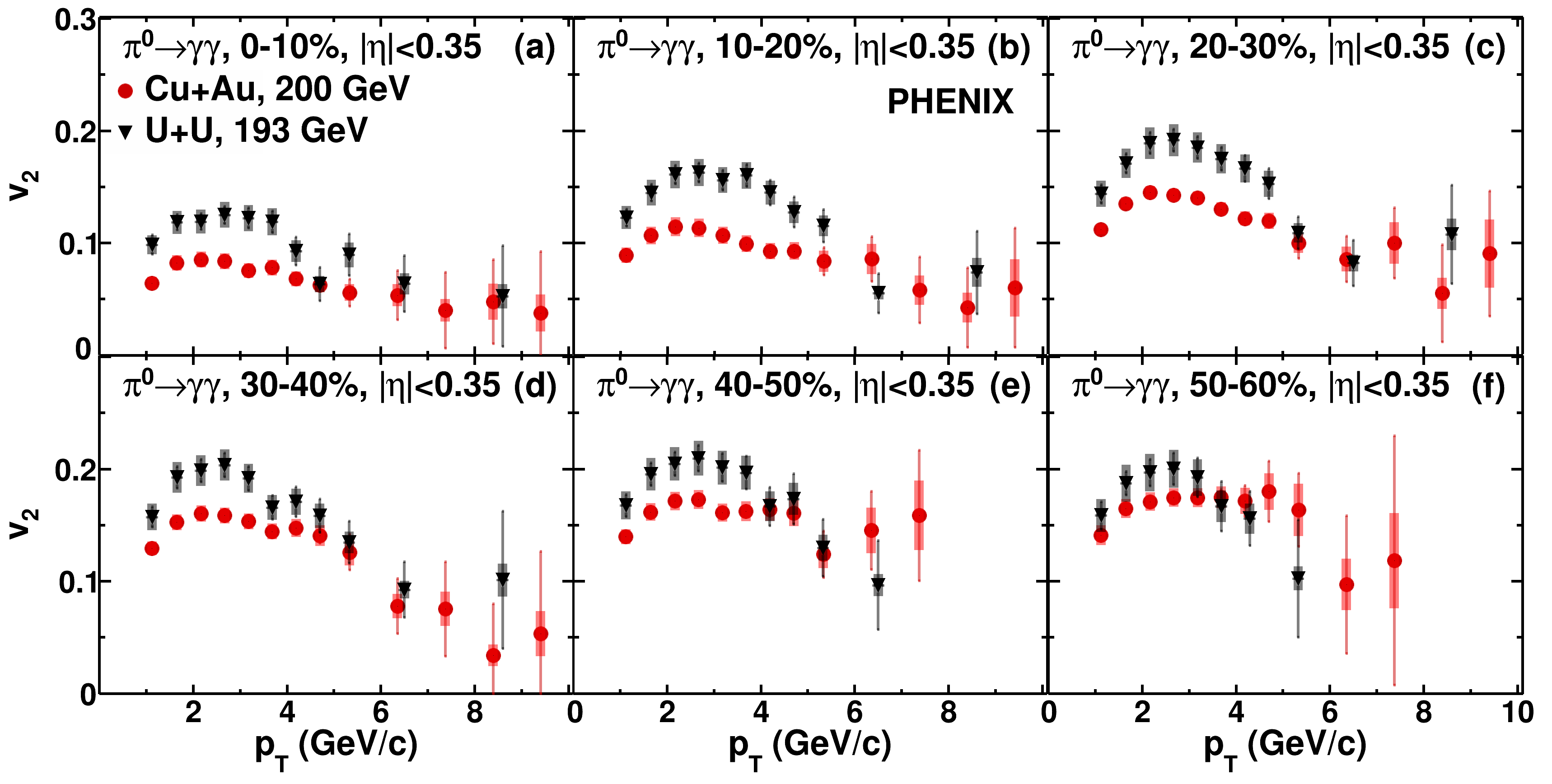}
\caption{The \pio \vtwo as a function of \pt measured at midrapidity in 
0\%--10\%, 10\%--20\%, 20\%--30\%, 30\%--40\%, 40\%--50\%, 50\%--60\% 
centrality bins of \cuau and \uu collisions. The error bars denote 
statistical uncertainties and the shaded boxes represent the systematic 
uncertainties.}
\label{fig:v2Obtained}
\end{figure*}

%%%%%%%%%%%%%%%%%%%%%%%%%%%%%%%%%%%%%%%%%%%%%%%%%%%%%%% Fig_5
\begin{figure}[]
\includegraphics[width=1.0\linewidth]{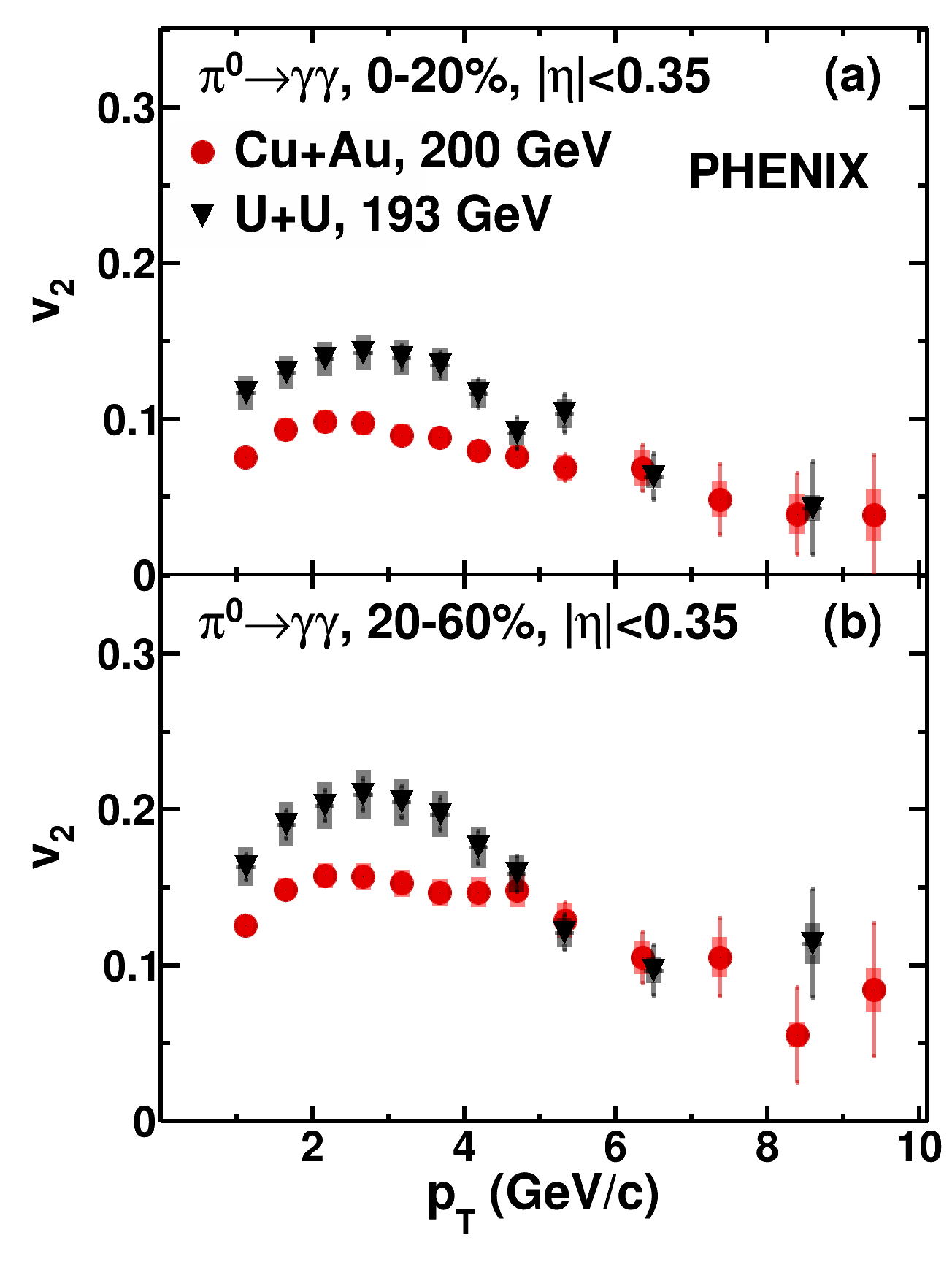}
\caption{The \pio \vtwo as a function of \pt measured at midrapidity in 
0\%--20\% and 20\%--60\% centrality bins of \cuau and \uu collisions. The 
error bars denote statistical uncertainties and the shaded boxes 
represent the systematic uncertainties.}
\label{fig:v2ObtainedMerged}
\end{figure}

\subsection{Azimuthal dependence of \pio nuclear-modification factors 
and associated fractional parton-energy losses}

The procedure for calculating azimuthal dependence of \pio suppression 
(\rabphi) and the associated fractional parton-energy losses (\slossphi) 
is based on previous publications~\cite{PRC_pi0_AUAU_2007, 
PRC_pi0_AUAU_2009, PRC_AUAU_PRODUCTION_2013} and uses data on the 
inclusive ($\varphi$-integrated) nuclear-modification factors of \pio 
(\rabincl) in \cuau and \uu collisions~\cite{ZHARKO,RADZEVICH} and the 
invariant spectrum of \pio in \pp collisions~\cite{pp}. 

At lower $p_T$, the $v_2$ primarily originates from elliptic flow.  At 
higher $p_T$, the $v_2$ primarily originates from the path-length 
dependence of parton-energy loss in the medium. The azimuthal 
dependence of the \pio nuclear-modification factors is determined from 
the measured $v_2$ values using
%\begin{linenomath}
\begin{equation}
    \label{eq:TOFm2}
    \rabphi = \rabincl(1+2v_{2}(\pt)\cos[2\deltaphi]).
\end{equation}
%\end{linenomath}

The suppression of \pio mesons at high transverse momentum, $p_T>5$ 
GeV/$c$, can be interpreted as a consequence of jet 
quenching~\cite{rabJetQuench}. Therefore, the associated effective 
fractional parton-energy loss can be estimated~\cite{QGP2, 
PRC_pi0_AUAU_2007} from the \pio nuclear-modification factor (\rabincl) 
as
%\begin{linenomath}
\begin{equation}
    \label{eq:slossincl}
    \slossincl = 1-R_{AB}^{1/(\beta-2)}(p_T),
\end{equation}
%\end{linenomath}
where $\beta$ is a power-law exponent obtained from fitting the 
invariant spectra of \pio in \pp collisions~\cite{pp}. Furthermore, the 
azimuthal dependence of fractional parton-energy losses can be 
calculated~\cite{PRC_pi0_AUAU_2007} from the obtained \rabphi as
%\begin{linenomath}
\begin{equation}
    \label{eq:slossphi}
    \slossphi = 1-R_{AB}^{1/(\beta-2)}(\Delta\varphi,p_T).
\end{equation}
%\end{linenomath}

The calculation of \slossphi values is based on the \rabphi values and a 
power-law approximation of the \pp spectrum. The \slossphi values should 
therefore be interpreted as an effective fractional momentum shift, or 
energy-loss proxy, rather than a direct event-by-event measurement of 
parton-energy loss~\cite{PRC_pi0_AUAU_2007,PRC_AUAU_PRODUCTION_2013}.

\section{SYSTEMATIC UNCERTAINTIES}

The dominant sources of systematic uncertainty in the measurement of 
\pio \vtwo are:

\begin{itemize}

\item[1] {{\bf The energy asymmetry of the photon pair.} 
To evaluate the effect of the energy asymmetry cut, \vtwo measurements 
were performed with $|\alpha|<0.6$ and without an energy-asymmetry cut 
and then compared with the data obtained using the cut $|\alpha|<0.8$. The 
uncertainties of $\approx2\%$ and $\approx3\%$ were determined in \cuau 
and \uu collision systems, respectively.}

\item[2] {{\bf The energy of photon candidates.} 
The uncertainty obtained from this source was estimated by comparing 
\vtwo measurements using three photon-energy ranges: $E_{\gamma}>0.4$ 
GeV (the default cut), $E_{\gamma}>0.3$ GeV, and $E_{\gamma}>0.5$ GeV. 
This uncertainty depends on the value of \pt: it decreases with 
increasing \pt and reaches 5\% at the lowest \pt value.}

\item[3] {{\bf Shower-shape cut (photon identification cut)~\cite{EMCal}.}
This cut helps to reduce the contribution from nonphotonic particles 
registered in the EMCal. Estimation of systematic uncertainty from this 
source is carried out by turning off this cut. This leads to a 
systematic uncertainty in \vtwo measurements of up to $\approx4\%$ in 
\cuau collisions and up to $\approx6\%$ in \uu collisions.}

\item[4] {{\bf Detector for the event-plane measurement.}
Independent measurements of integrated \vtwo values were compared using 
different event-plane detectors: the FVTX and the BBC for the \cuau 
analysis; and the MPC and the BBC for the \uu analysis. An uncertainty 
of $<4$\%--7\% that slightly varies with centrality was observed.}

\item[5] {{\bf Extraction of raw \pio yields.} To calculate systematic 
uncertainties from this source, several different functions are used to 
approximate the residual background combined with several different 
integration windows to estimate the \pio signal.} The obtained 
systematic uncertainties increase with increasing \pt and reach maximum 
values of $\approx40\%$ and $\approx20\%$ at the highest \pt in \cuau 
and \uu collisions, respectively. Table~\ref{table:syst} summarizes the 
systematic uncertainties for the measurements of \pio \vtwo, which are 
categorized by the following types:

\begin{itemize}

\item[{\bf A}]{\pt-uncorrelated point-to-point uncertainties (which are 
dominated by the statistical precision of the data);}

\item[{\bf B}]{\pt-correlated point-to-point uncertainties;}

\item[{\bf C}]{global normalization uncertainties.}

\end{itemize}

\end{itemize}

%========================================================== Table_II
\begin{table*}[tb]
\caption{Systematic uncertainties for \pio \vtwo values at different \pt 
measured in \cuau collisions at \sqsn~=~200 GeV and \uu at \sqsn~=~193 
GeV. The type of uncertainties are described in the text. Values with a 
range indicate the variation of the uncertainty over the different 
centrality bins.}
\label{table:syst}
\begin{ruledtabular}
\begin{tabular}{cccccc} % Changed to 6 columns
Source & Type & \multicolumn{2}{c}{\cuau} & \multicolumn{2}{c}{\uu} \\
 \pt &  & 2.75 GeV/$c$ & 5.5 GeV/$c$ & 2.75 GeV/$c$ & 5.5 GeV/$c$ \\
\hline
Energy asymmetry ($\alpha$) & C &  2.0\%  & 2.0\% & 3.3\%  & 3.3\% \\
Energy of photon ($E_{\gamma}$) & B & 0.5\%--1\% & 0.1\%--0.5\% & 1.5\%--2.5\% & 0.1\%--0.5\% \\
Shower shape & C & 2.0\%--4.0\% & 2.0\%--4.0\% & 4.0\%--6.5\% & 4.0\%--6.5\% \\
Event-plane detector & C & 4.0\%--7.0\% & 4.0\%--7.0\% & 2.0\%--3.0\% & 2.0\%--3.0\% \\
Raw \pio yield & B & 1.0\%--3.0\% & 8.0\%--14.0\% & 1.0\%--3.0\% & 5.0\%--13.0\% \\
\end{tabular}
\end{ruledtabular}
\end{table*}

\section{RESULTS}

\subsection{Elliptic flow}

%%%%%%%%%%%%%%%%%%%%%%%%%%%%%%%%%%%%%%%%%%%%%%%%%%%%%%% Fig_6
\begin{figure*}[]
\includegraphics[width=0.99\linewidth]{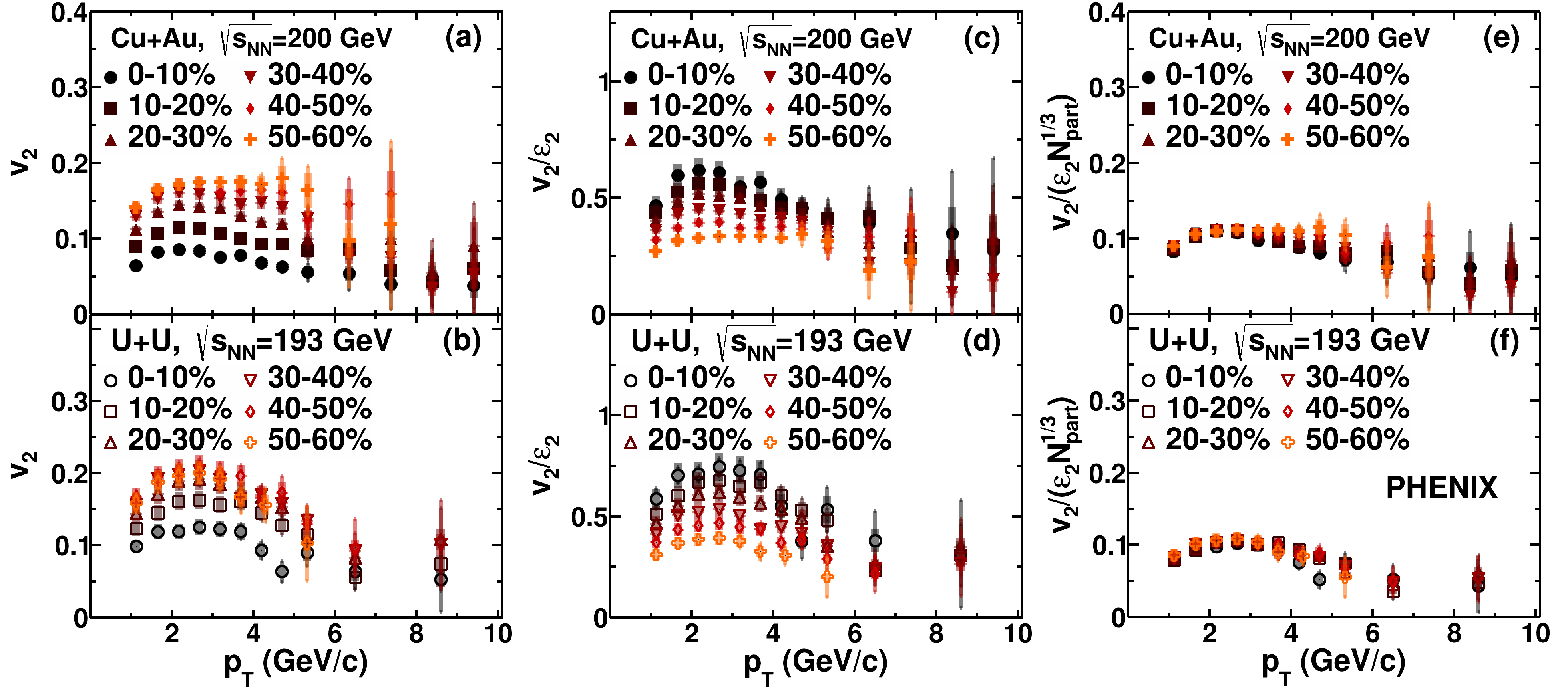}
\caption{The comparison of \pio (a, b) \vtwo, (c, d) 
\vtwo$/\varepsilon_{2}$, (e, f) \vtwo$/(\hydroscaling)$ measured as a 
function of \pt in \cuau and \uu collisions at midrapidity 
($|\eta|<0.35$) at 200 and 193 GeV, respectively. The uncertainties of 
$\varepsilon_{2}$ and \Npart are included in the total systematic 
uncertainties of $\vtwo(\pt)/\varepsilon_{2}$ and 
$\vtwo(\pt)/(\hydroscaling)$ values.}
\label{fig:v2All10}
\end{figure*}

%%%%%%%%%%%%%%%%%%%%%%%%%%%%%%%%%%%%%%%%%%%%%%%%%%%%%%% Fig_7
\begin{figure*}[]
\includegraphics[width=0.99\linewidth]{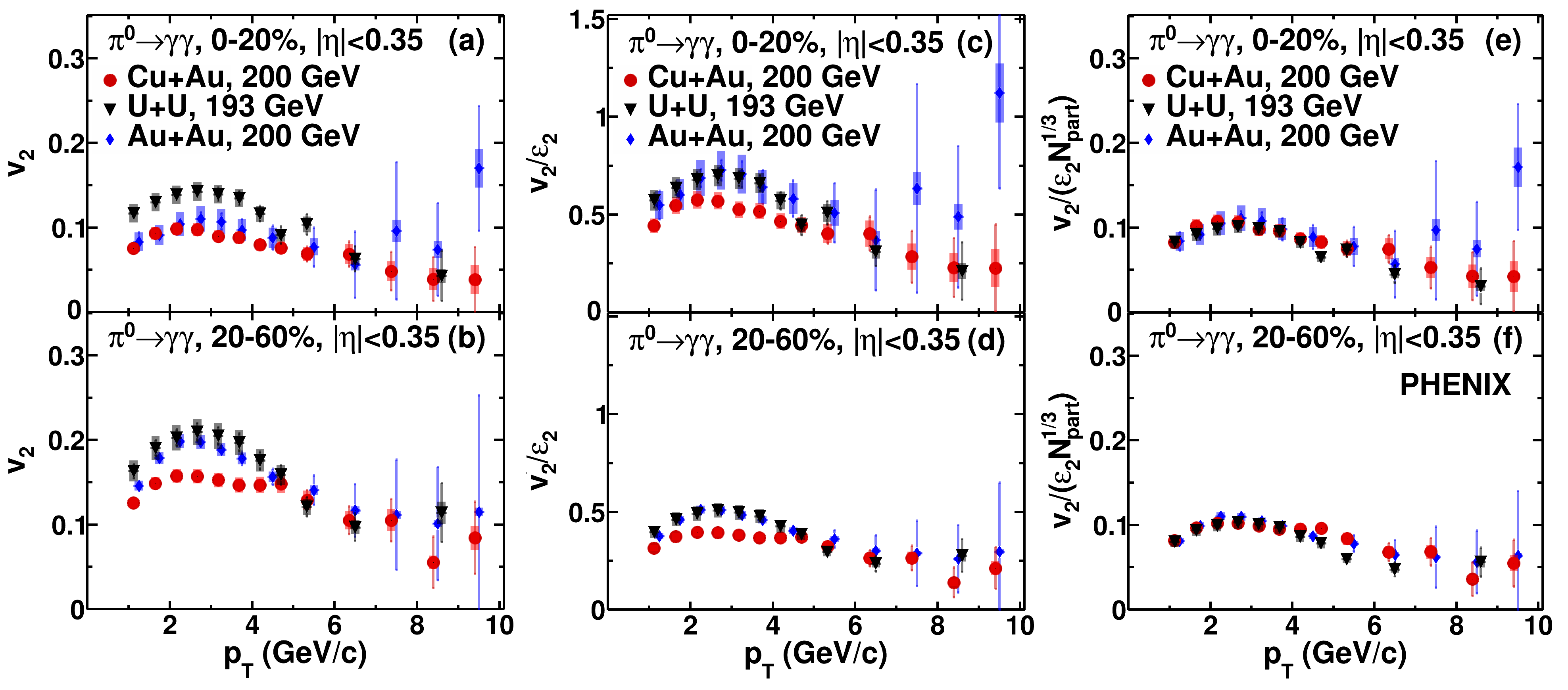}
\caption{The comparison of \pio (a, b) \vtwo, (c, d) 
$\vtwo/\varepsilon_{2}$ and (e, f) $\vtwo/(\hydroscaling)$ measured as a 
function of \pt in \cuau, \auau~\cite{PRC_pi0_AUAU_2010, PPG129} and \uu 
collisions at midrapidity ($|\eta|<0.35$) at 200 and 193 GeV, 
respectively. The uncertainties of $\varepsilon_{2}$ and \Npart are 
included in the total systematic uncertainties of 
$\vtwo(\pt)/\varepsilon_{2}$ and $\vtwo(\pt)/(\hydroscaling)$ values.}
\label{fig:v2sDiffSystems}
\end{figure*}

%%%%%%%%%%%%%%%%%%%%%%%%%%%%%%%%%%%%%%%%%%%%%%%%%%%%%%% Fig_8
\begin{figure*}[]
\includegraphics[width=0.99\linewidth]{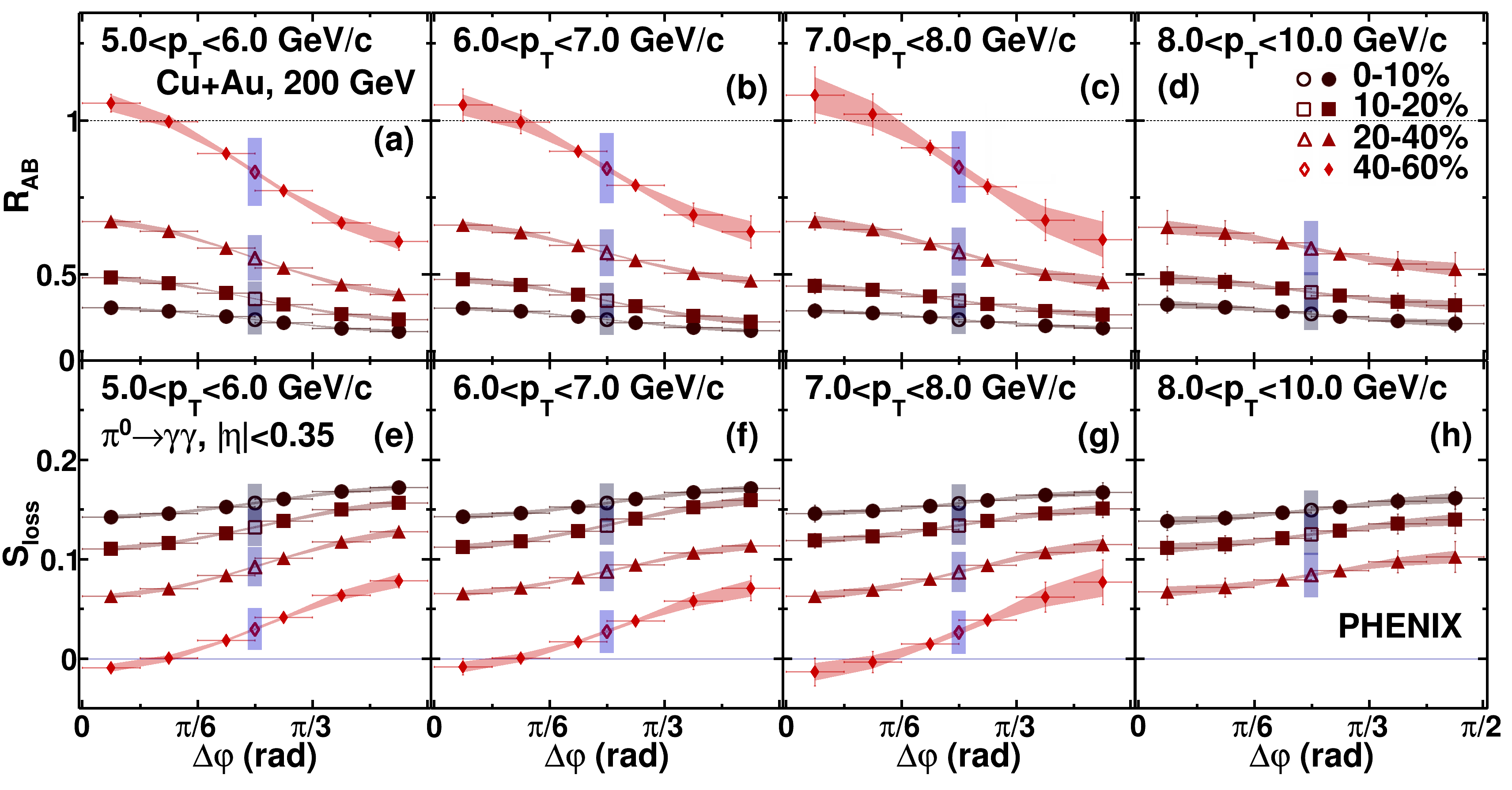}
\caption{The comparison of \pio (a, b, c, d) \rabphi and (e, f, g, h) 
associated \slossphi in \cuau at 200 GeV. Closed symbols represent the 
$\Delta\varphi$-dependent values, and open symbols indicate inclusive 
data taken from Ref.~\cite{ZHARKO}. The inclusive values of \rabincl and 
\slossincl located at $\Delta\varphi=\pi/4$ are a visual convention 
representing the $\varphi$-integrated values. The filled areas represent 
the systematic uncertainties of \rabphi and \slossphi values, and the 
shaded rectangles show the systematic uncertainties of \rabincl and 
\slossincl values.}
\label{fig:sloss_cuau}
\end{figure*}

%%%%%%%%%%%%%%%%%%%%%%%%%%%%%%%%%%%%%%%%%%%%%%%%%%%%%%% Fig_9
\begin{figure*}[]
\includegraphics[width=0.99\linewidth]{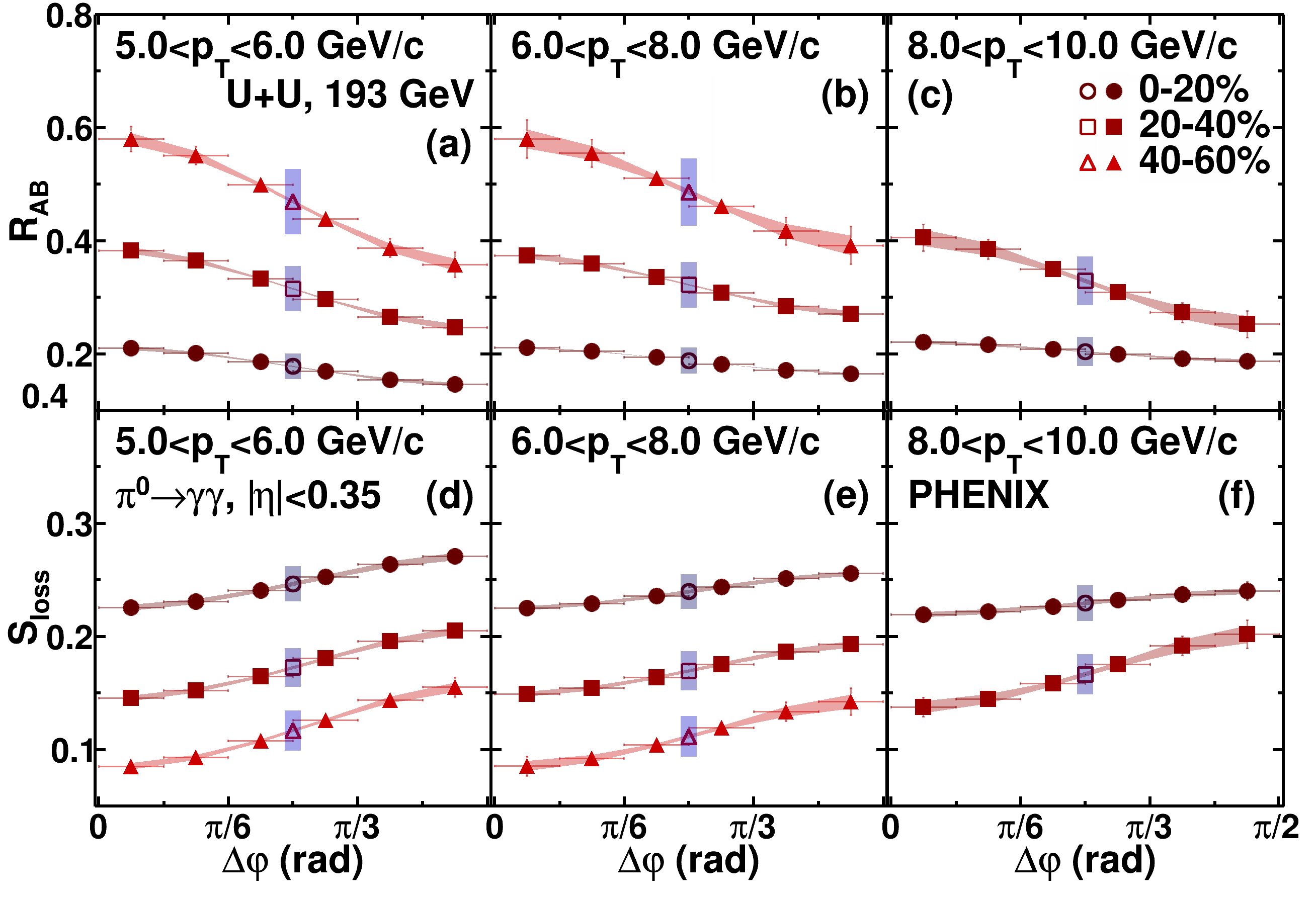}
\caption{The comparison of \pio (a, b, c) \rabphi and (d, e, f) 
associated \slossphi in \uu at 193 GeV. Closed symbols represent the 
$\Delta\varphi$-dependent values, and open symbols indicate inclusive 
data taken from Ref.~\cite{RADZEVICH}. The inclusive values of \rabincl 
and \slossincl located at $\Delta\varphi=\pi/4$ are a visual convention 
representing the $\varphi$-integrated values. The filled areas represent 
the systematic uncertainties of \rabphi and \slossphi values, and the 
shaded rectangles show the systematic uncertainties of \rabincl and 
\slossincl values.}
\label{fig:sloss_uu}
\end{figure*}

%%%%%%%%%%%%%%%%%%%%%%%%%%%%%%%%%%%%%%%%%%%%%%%%%%%%%%% Fig_10
\begin{figure}[]
\includegraphics[width=1.0\linewidth]{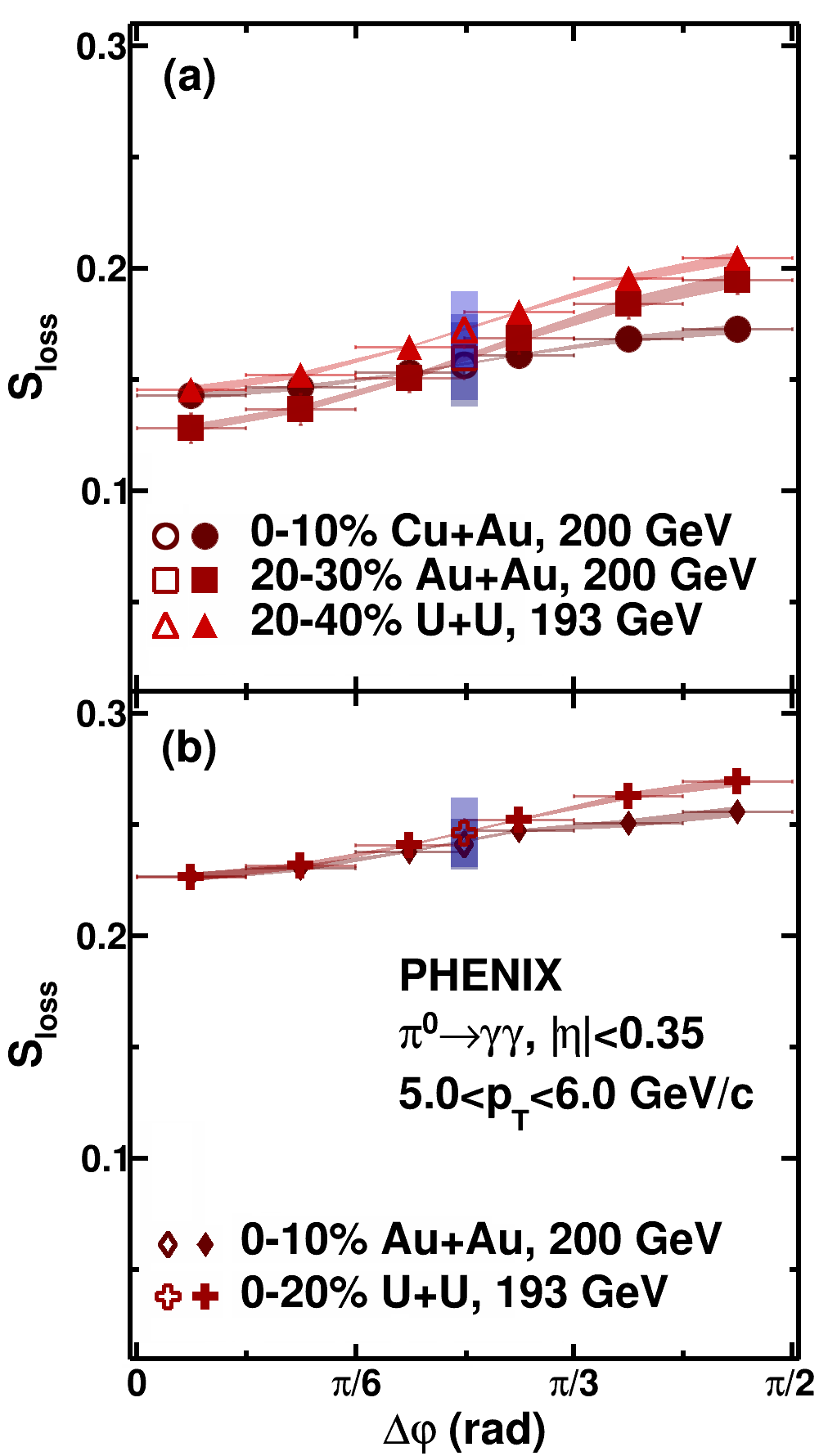}
\caption{The comparison of \slossphi for $5$ GeV/$c$ $<\pt<6$ GeV/$c$ 
in \cuau, \auau at 200 GeV and \uu at 193 GeV for two values of 
nucleon-participant number: (a) $N_{\rm part}\approx170$ and (b) 
$N_{\rm part}\approx330$. The inclusive values of \slossincl located at 
$\Delta\varphi=\pi/4$ are a visual convention representing the 
$\varphi$-integrated values. The filled areas represent the systematic 
uncertainties of \slossphi values, and the shaded rectangles show the 
systematic uncertainties of \slossincl values.} 
\label{fig:sloss_cuau_uu} 
\end{figure}

Figures~\ref{fig:v2Obtained} and~\ref{fig:v2ObtainedMerged} show the 
\pio-meson \vtwo as a function of \pt obtained in six centrality bins 
(0\%--10\%, 10\%--20\%, 20\%--30\%, 30\%--40\%, 40\%--50\%, 50\%--60\%) 
and 2 merged centrality bins (0\%--20\% and 20\%--60\%) of \cuau and 
\uu collisions. The obtained \vtwo values increase when moving toward 
more peripheral collisions, starting from the most central collisions 
of 0\%--10\% to the 30\%--40\% centrality bin in both collision 
systems. In more peripheral collisions (40\%--50\%, 50\%--60\%) the 
measured elliptic flow values are consistent within the uncertainties 
with those obtained in the centrality bin of 30\%--40\%. The increase 
of \vtwo from central to midcentral collisions reflects the increasing 
ellipticity of the nuclear-overlap region. Also, the 
system size becomes smaller, which has a countervailing effect on 
\vtwo, hence \vtwo does not continue to grow in even more peripheral 
collisions. For all centrality bins in both collision systems, the 
elliptic flow values increase to $\approx3$ GeV/$c$ and then tend to 
decrease. This trend holds except for the most-peripheral collisions 
(50\%--60\%) of \cuau, where, due to the lower multiplicity, nonflow 
effects may play a more significant role. For large values of 
transverse momentum ($\pt>5$ GeV/$c$) the magnitude of the obtained 
\vtwo is nonzero. This finding of $\vtwo(\pt)$ behavior reflects a 
change in the dominant mechanism from the hydrodynamic expansion of QGP 
at low $\pt\leq3$ GeV/$c$ to the path-length-dependent jet quenching of 
the leading \pio at high $\pt>5$ GeV/$c$, which is consistent with 
previous works devoted to the \vtwo measurements of \pio in Au$+$Au 
collisions at 200 
GeV~\cite{PRC_pi0_AUAU_2007,PRC_pi0_AUAU_2009,PRC_pi0_AUAU_2010,PPG129}.

Figure~\ref{fig:v2All10} presents (a, b) $\vtwo(\pt)$, (c, d) 
$\vtwo(\pt)/\ecc$, and (e, f) $\vtwo(\pt)/(\hydroscaling)$ values 
for \pio mesons obtained in six centrality bins (0\%--10\%, 10\%--20\%, 
20\%--30\%, 30\%--40\%, 40\%--50\%, 50\%--60\%) of \cuau and \uu 
collisions. To investigate the dependence of the elliptic flow 
development on the initial geometry, the influence of the shape of the 
nuclear-overlap region in different centrality bins was studied by 
scaling the obtained \vtwo values with eccentricity (\ecc) values taken 
from Ref.~\cite{PPG183}. However, Fig.~\ref{fig:v2All10}(c, d) shows 
that such scaling does not lead to universal behavior. This is because 
it only accounts for the geometry of the collisions and does not take into 
account the system size. Because \Npart can be used to estimate the 
volume of QGP formed in heavy-ion collisions~\cite{PPG124}, \npart is 
taken to be an estimate of the radius. Both the geometry and the system 
size can be accounted for by scaling the \vtwo by the factor 
\hydroscaling, which is shown in Fig.~\ref{fig:v2All10}(e, f).  The 
$\vtwo(\pt)/\hydroscaling$ values are consistent within the 
uncertainties up to $\pt\approx4$ GeV/$c$ in all centrality bins of 
\cuau and \uu collisions.

For a more detailed study of the dependence of the \vtwo development on 
the initial collision geometry, a comparison of the obtained elliptic 
flow values in asymmetric and deformed collision systems with those in 
symmetric Au$+$Au collisions~\cite{PRC_pi0_AUAU_2010, PPG129} was made. 
The results of this comparison are presented in 
Fig.~\ref{fig:v2sDiffSystems}, which shows \pio-meson \vtwo, 
$\vtwo/\varepsilon_{2}$, and $\vtwo/(\hydroscaling)$ as a function of 
\pt in the centrality bins 0\%--20\%, 20\%--60\% in \cuau, \auau, and 
\uu collisions. The $\vtwo(\pt)/\hydroscaling$ values agree within the 
uncertainties up to $\pt\approx4$ GeV/$c$ in central and peripheral 
collisions of symmetric, asymmetric, and deformed nuclei.

\subsection{Azimuthal dependence of \pio nuclear-modification factors 
and associated fractional parton-energy losses}

Figure~\ref{fig:sloss_cuau} presents the nuclear-modification factors of 
\pio as a function of emission angle relative to the reaction plane 
for four centrality bins (0\%--10\%, 10\%--20\%, 20\%--40\%, 40--60\%) of 
\cuau collisions. The corresponding fractional parton-energy losses are 
also shown in Fig.~\ref{fig:sloss_cuau}. The measurements are for  
four \pt intervals in the range 5~GeV/$c<\pt<10$~GeV/$c$.

The \rabphi and \slossphi values calculated for \uu collisions in the 
\pt range 5~GeV/$c$ $<\pt<10$~GeV/$c$ are shown in 
Fig.~\ref{fig:sloss_uu}. For both collision systems, inclusive 
\slossincl values were calculated using Eq.~\ref{eq:slossincl}, where 
$\beta$ value and the invariant spectra of \pio in \pp collisions are 
taken from Ref.~\cite{pp}.

The systematic uncertainties of the $\Delta\varphi$-dependent values 
are calculated from the total systematic uncertainties of the measured 
\vtwo values (Type B), while for the inclusive data they are taken from 
Ref.~\cite{ZHARKO, RADZEVICH} (Type C). The Type B and C systematic 
uncertainties are presented by filled boxes. Error bars in 
Fig.~\ref{fig:sloss_cuau} and Fig.~\ref{fig:sloss_uu} for all 
measurements denote combined statistical uncertainties and Type A 
systematic uncertainties.

The calculated values of \rabphi reach their maximum when aligned with 
the reaction plane and their minimum when perpendicular to it. The 
suppression of \pio and the associated fractional parton-energy losses 
depend not only on the size of nuclear-overlap region, but also on its 
shape.  The difference between the maximum and minimum values of \rabphi 
and \slossphi increases when moving toward more peripheral collision, 
which is consistent with the path-length dependence of energy loss of 
partons passing through the QGP. In addition, the measured 
$\Delta\varphi$-dependent values in different \pt intervals agree within 
the uncertainties, indicating the same \pio suppression pattern in the 
presented \pt ranges.

Figure~\ref{fig:sloss_cuau_uu} shows a comparison of \slossphi for $5$ 
GeV/$c$ $<\pt<6$ GeV/$c$ in \cuau, \auau, and \uu collisions for two 
values of the nucleon-participant number, $N_{\rm part}\approx170$ and 
$N_{\rm part}\approx330$. The azimuthal dependence of the fractional 
parton-energy losses in \auau collisions was determined by 
Eq.~\ref{eq:slossphi} using the obtained \rabphi from 
Ref.~\cite{PRC_AUAU_PRODUCTION_2013}. The inclusive values of 
\slossincl located at $\Delta\varphi=\pi/4$ in \cuau, \auau, and \uu 
collisions are consistent with each other within the uncertainties. The 
values of \slossphi exhibit slightly different behavior, which may be 
due to different ellipticities of the nuclear-overlap regions. This 
difference is most evident at $\Npart\approx170$ between \uu, \auau and 
\cuau data. Ellipticities of the nuclear-overlap regions can be 
quantified by $\varepsilon_{2}$ values. Table~\ref{table:data} presents 
values of $\Npart$ and $\varepsilon_{2}$ for different centrality bins 
in \cuau, \auau collisions at \sqsn~=~200 GeV and \uu collisions at 
\sqsn~=~193 GeV taken from Ref.~\cite{PPG183},~\cite{Masui:2009qk}. The 
maximum difference in $\varepsilon_{2}$ values is observed at 
$\Npart\approx170$ when comparing the values for \uu and \auau 
collisions to \cuau collisions. At $\Npart\approx330$, the discrepancy 
in $\varepsilon_{2}$ values for \uu and \auau collisions is also 
significant. These comparisons of $\varepsilon_{2}$ values agree with 
observed \slossphi dependencies. Thus, possible differences in 
modulations of $\Delta\varphi$-dependent values between collision 
systems at approximately the same \Npart values may reflect the 
different eccentricities of the selected centrality bins.

%========================================================== Table_III
\begin{table}[tb]
\caption{Values of $\Npart$ and $\varepsilon_{2}$ for different 
centrality bins in \cuau, \auau collisions at \sqsn~=~200 GeV and \uu 
collisions at \sqsn~=~193 GeV~\cite{PPG183},~\cite{Masui:2009qk}.}
\label{table:data}
\begin{ruledtabular}
\begin{tabular}{cccc}
Coll. & Centrality & \Npart & $\varepsilon_{2}$\\
\hline
\cuau &  0\%--10\% & $177.2\pm5.2$    & $0.138\pm0.011$ \\
\auau &  0\%--10\% & $325.2\pm3.3$    & $0.103\pm0.003$ \\
      & 20\%--30\% & $166.6\pm5.4$    & $0.284\pm0.006$ \\
\uu   &  0\%--20\% & $334.5\pm11.5$   & $0.204\pm0.011$ \\
      & 20\%--40\% & $168.0\pm14.0$   & $0.345\pm0.030$ \\
\end{tabular}
\end{ruledtabular}
\end{table}

% & & $\pm5.2$ & $\pm0.011$ \\
%& & $\pm3.3$ & $\pm0.003$ \\
%& & $\pm5.4$ & $\pm0.006$ \
%& & $\pm11.5$ & $\pm0.011$ \\
%& & $\pm14.0$ & $\pm0.030$ \\

%%%%%%%%%%%%%%%%%%%%%%%%%%%%%%%%%%%%%%%%%%%%%%%%%%%%%%%%%%%%%%%%%%%%%%%%%%%%%%

\section{SUMMARY}

The PHENIX experiment has measured the \vtwo of \pio mesons as a 
function of \pt in \cuau and \uu collisions at \sqsn~=~200 GeV and 
\sqsn~=~193 GeV, respectively. The measurements were performed in a wide 
range of \pt up to $\approx10$ GeV/$c$ in six centrality bins (0\%--10\%, 
10\%--20\%, 20\%--30\%, 30\%--40\%, 40\%--50\%, 50\%--60\%). Using the 
measured \vtwo values at high $\pt>5$ GeV/$c$, the azimuthal dependence 
of the \pio suppression \rabphi and the associated fractional 
parton-energy losses \slossphi have been estimated in four centrality bins of 
\cuau collisions (0\%--10\%, 10\%--20\%, 20\%--40\%, 40\%--60\%) and in 
three centrality bins (0\%--20\%, 20\%--40\%, 40\%--60\%) of the \uu 
collision system. Comparisons of the measured \vtwo and \slossphi values 
with previously obtained PHENIX results in \auau collisions at 200 GeV 
have been provided.

The observed trend of $\vtwo(\pt)$ in the asymmetric \cuau collision 
system and in collisions of deformed nuclei \uu is the same as in the 
symmetric \auau collision system~\cite{PRC_pi0_AUAU_2009}, and can be 
described by a change in the dominant mechanism of \vtwo development 
from the collective hydrodynamic expansion of the QGP at $\pt\leq3$ 
GeV/$c$ to the in-medium parton-energy loss at $\pt>5$ GeV/$c$. To 
further clarify these two mechanisms of \vtwo generation:  
the \vtwo was scaled by a factor \hydroscaling and the 
$\Delta\varphi$-dependent nuclear-modification factors of \pio and the 
associated fractional parton-energy losses were calculated. The 
$\vtwo(\pt)/\hydroscaling$ values are consistent within the 
uncertainties up to $\pt\approx4$~GeV/$c$ in all centrality bins of 
\cuau, \auau, and \uu collisions, which can be interpreted as the 
predominance of elliptic flow~\cite{Taranenko}. The calculated values of 
\rabphi and \slossphi in \cuau and \uu collisions have the same pattern 
as in previous measurements in \auau collisions. All values vary 
proportionally to the shape of the created QGP (characterized by the 
$\varepsilon_{2}$) and its volume (characterized by \Npart). Moreover, 
obtained \slossincl in \cuau, \auau, \uu collision systems for 
comparable \Npart are consistent within uncertainties. However, 
modulations of \slossphi values across all three collision systems at 
approximately the same \Npart have slight differences which can be 
described by different eccentricities of nuclear-overlap regions. This 
provides additional sensitivity to the path-length and medium-density 
dependence of parton-energy losses. These observations support a common 
interpretation of the \pio \vtwo behavior across different initial 
geometries of heavy-ion collisions: hydrodynamic expansion of QGP as a 
main driver of \vtwo generation up to $\pt\approx4$ GeV/$c$ and 
path-length-dependent parton-energy losses as a source of \vtwo at 
$\pt>5$ GeV/$c$.

%%%%%%%%%%%%%%%%%%%%%%  ACKNOWLEDGMENTS}  %%%%% MGS22a version
%% 2018 change in Korea
%%% 2021 change in dropping Brazil, Germany, and Pakistan, because
%%%      they no longer have active MGS and left PHENIX before 2015
%% 2024 add HUN-REN ATOMKI [and remove some] (Hungary)

\section*{ACKNOWLEDGMENTS}

We thank the staff of the Collider-Accelerator and Physics
Departments at Brookhaven National Laboratory and the staff of
the other PHENIX participating institutions for their vital
contributions.  
We acknowledge support from the Office of Nuclear Physics in the
Office of Science of the Department of Energy,
the National Science Foundation,
Abilene Christian University Research Council,
Research Foundation of SUNY, and
Dean of the College of Arts and Sciences, Vanderbilt University
(U.S.A),
Ministry of Education, Culture, Sports, Science, and Technology
and the Japan Society for the Promotion of Science (Japan),
Conselho Nacional de Desenvolvimento Cient\'{\i}fico e
Tecnol{\'o}gico and Funda\c c{\~a}o de Amparo {\`a} Pesquisa do
Estado de S{\~a}o Paulo (Brazil),
Natural Science Foundation of China (People's Republic of China),
Croatian Science Foundation and
Ministry of Science and Education (Croatia),
Ministry of Education, Youth and Sports (Czech Republic),
Centre National de la Recherche Scientifique, Commissariat
{\`a} l'{\'E}nergie Atomique, and Institut National de Physique
Nucl{\'e}aire et de Physique des Particules (France),
Bundesministerium f\"ur Bildung und Forschung, Deutscher
Akademischer Austausch Dienst, and Alexander von Humboldt Stiftung (Germany),
J. Bolyai Research Scholarship, EFOP, HUN-REN ATOMKI, NKFIH,
MATE KKF, and OTKA (Hungary), 
Department of Atomic Energy and Department of Science and Technology (India),
Israel Science Foundation (Israel),
Basic Science Research and SRC(CENuM) Programs through NRF
funded by the Ministry of Education and the Ministry of
Science and ICT (Korea).
Physics Department, Lahore University of Management Sciences (Pakistan),
Ministry of Education and Science, Russian Academy of Sciences,
Federal Agency of Atomic Energy (Russia),
VR and Wallenberg Foundation (Sweden),
University of Zambia, the Government of the Republic of Zambia (Zambia),
the U.S. Civilian Research and Development Foundation for the
Independent States of the Former Soviet Union,
the Hungarian American Enterprise Scholarship Fund,
the US-Hungarian Fulbright Foundation,
and the US-Israel Binational Science Foundation.

\section*{DATA AVAILABILITY}

The data that support the findings of this article are not publicly 
available. The values in the plots associated with this article are 
stored in HEPData~\cite{hepdata}.

%\bibliography{ppg260x0}

\begin{thebibliography}{33}%
\makeatletter
\providecommand \@ifxundefined [1]{%
 \@ifx{#1\undefined}
}%
\providecommand \@ifnum [1]{%
 \ifnum #1\expandafter \@firstoftwo
 \else \expandafter \@secondoftwo
 \fi
}%
\providecommand \@ifx [1]{%
 \ifx #1\expandafter \@firstoftwo
 \else \expandafter \@secondoftwo
 \fi
}%
\providecommand \natexlab [1]{#1}%
\providecommand \enquote  [1]{``#1''}%
\providecommand \bibnamefont  [1]{#1}%
\providecommand \bibfnamefont [1]{#1}%
\providecommand \citenamefont [1]{#1}%
\providecommand \href@noop [0]{\@secondoftwo}%
\providecommand \href [0]{\begingroup \@sanitize@url \@href}%
\providecommand \@href[1]{\@@startlink{#1}\@@href}%
\providecommand \@@href[1]{\endgroup#1\@@endlink}%
\providecommand \@sanitize@url [0]{\catcode `\\12\catcode `\$12\catcode
  `\&12\catcode `\#12\catcode `\^12\catcode `\_12\catcode `\%12\relax}%
\providecommand \@@startlink[1]{}%
\providecommand \@@endlink[0]{}%
\providecommand \url  [0]{\begingroup\@sanitize@url \@url }%
\providecommand \@url [1]{\endgroup\@href {#1}{\urlprefix }}%
\providecommand \urlprefix  [0]{URL }%
\providecommand \Eprint [0]{\href }%
\providecommand \doibase [0]{https://doi.org/}%
\providecommand \selectlanguage [0]{\@gobble}%
\providecommand \bibinfo  [0]{\@secondoftwo}%
\providecommand \bibfield  [0]{\@secondoftwo}%
\providecommand \translation [1]{[#1]}%
\providecommand \BibitemOpen [0]{}%
\providecommand \bibitemStop [0]{}%
\providecommand \bibitemNoStop [0]{.\EOS\space}%
\providecommand \EOS [0]{\spacefactor3000\relax}%
\providecommand \BibitemShut  [1]{\csname bibitem#1\endcsname}%
\let\auto@bib@innerbib\@empty
%</preamble>
\bibitem [{\citenamefont {Shuryak}(1980)}]{QGP1}%
  \BibitemOpen
  \bibfield  {author} {\bibinfo {author} {\bibfnamefont {E.~V.}\ \bibnamefont
  {Shuryak}},\ }\bibfield  {title} {\bibinfo {title} {Quantum chromodynamics
  and the theory of superdense matter},\ }\href
  {https://doi.org/https://doi.org/10.1016/0370-1573(80)90105-2} {\bibfield
  {journal} {\bibinfo  {journal} {Phys. Rept.}\ }\textbf {\bibinfo {volume}
  {61}},\ \bibinfo {pages} {71} (\bibinfo {year} {1980})}\BibitemShut {NoStop}%
\bibitem [{\citenamefont {Adcox}\ \emph {et~al.}(2005)\citenamefont {Adcox}
  \emph {et~al.}}]{QGP2}%
  \BibitemOpen
  \bibfield  {author} {\bibinfo {author} {\bibfnamefont {K.}~\bibnamefont
  {Adcox}} \emph {et~al.} (\bibinfo {collaboration} {PHENIX Collaboration}),\
  }\bibfield  {title} {\bibinfo {title} {{Formation of dense partonic matter in
  relativistic nucleus+nucleus collisions at {RHIC}: Experimental evaluation by
  the PHENIX Collaboration}},\ }\href
  {https://doi.org/10.1016/j.nuclphysa.2005.03.086} {\bibfield  {journal}
  {\bibinfo  {journal} {Nucl. Phys. A}\ }\textbf {\bibinfo {volume} {757}},\
  \bibinfo {pages} {184–283} (\bibinfo {year} {2005})}\BibitemShut {NoStop}%
\bibitem [{\citenamefont {Collins}\ and\ \citenamefont
  {Perry}(1975)}]{AzimuthalAnisotropy}%
  \BibitemOpen
  \bibfield  {author} {\bibinfo {author} {\bibfnamefont {J.~C.}\ \bibnamefont
  {Collins}}\ and\ \bibinfo {author} {\bibfnamefont {M.~J.}\ \bibnamefont
  {Perry}},\ }\bibfield  {title} {\bibinfo {title} {{Superdense Matter:
  Neutrons Or Asymptotically Free Quarks?}},\ }\href
  {https://doi.org/10.1103/PhysRevLett.34.1353} {\bibfield  {journal} {\bibinfo
   {journal} {Phys. Rev. Lett.}\ }\textbf {\bibinfo {volume} {34}},\ \bibinfo
  {pages} {1353} (\bibinfo {year} {1975})}\BibitemShut {NoStop}%
\bibitem [{\citenamefont {Voloshin}\ \emph {et~al.}(2010)\citenamefont
  {Voloshin}, \citenamefont {Poskanzer},\ and\ \citenamefont
  {Snellings}}]{CollectivePhenomena}%
  \BibitemOpen
  \bibfield  {author} {\bibinfo {author} {\bibfnamefont {S.~A.}\ \bibnamefont
  {Voloshin}}, \bibinfo {author} {\bibfnamefont {A.~M.}\ \bibnamefont
  {Poskanzer}},\ and\ \bibinfo {author} {\bibfnamefont {R.}~\bibnamefont
  {Snellings}},\ }\bibfield  {title} {\bibinfo {title} {{Collective phenomena
  in noncentral nuclear collisions}},\ }\href
  {https://doi.org/10.1007/978-3-642-01539-7_10} {\bibfield  {journal}
  {\bibinfo  {journal} {Landolt-Bornstein}\ }\textbf {\bibinfo {volume} {23}},\
  \bibinfo {pages} {293} (\bibinfo {year} {2010})}\BibitemShut {NoStop}%
\bibitem [{\citenamefont {Heinz}\ and\ \citenamefont
  {Snellings}(2013)}]{Heinz:2013th}%
  \BibitemOpen
  \bibfield  {author} {\bibinfo {author} {\bibfnamefont {U.}~\bibnamefont
  {Heinz}}\ and\ \bibinfo {author} {\bibfnamefont {R.}~\bibnamefont
  {Snellings}},\ }\bibfield  {title} {\bibinfo {title} {{Collective flow and
  viscosity in relativistic heavy-ion collisions}},\ }\href
  {https://doi.org/10.1146/annurev-nucl-102212-170540} {\bibfield  {journal}
  {\bibinfo  {journal} {Ann. Rev. Nucl. Part. Sci.}\ }\textbf {\bibinfo
  {volume} {63}},\ \bibinfo {pages} {123} (\bibinfo {year} {2013})}\BibitemShut
  {NoStop}%
\bibitem [{\citenamefont {Gyulassy}\ \emph {et~al.}(2001)\citenamefont
  {Gyulassy}, \citenamefont {Vitev},\ and\ \citenamefont
  {Wang}}]{Gyulassy:2000gk}%
  \BibitemOpen
  \bibfield  {author} {\bibinfo {author} {\bibfnamefont {M.}~\bibnamefont
  {Gyulassy}}, \bibinfo {author} {\bibfnamefont {I.}~\bibnamefont {Vitev}},\
  and\ \bibinfo {author} {\bibfnamefont {X.-N.}\ \bibnamefont {Wang}},\
  }\bibfield  {title} {\bibinfo {title} {{High ${p_{T}}$ Azimuthal Asymmetry in
  Noncentral A+A at RHIC}},\ }\href
  {https://doi.org/10.1103/PhysRevLett.86.2537} {\bibfield  {journal} {\bibinfo
   {journal} {Phys. Rev. Lett.}\ }\textbf {\bibinfo {volume} {86}},\ \bibinfo
  {pages} {2537} (\bibinfo {year} {2001})}\BibitemShut {NoStop}%
\bibitem [{\citenamefont {Adare}\ \emph {et~al.}(2015)\citenamefont {Adare}
  \emph {et~al.}}]{PPG124}%
  \BibitemOpen
  \bibfield  {author} {\bibinfo {author} {\bibfnamefont {A.}~\bibnamefont
  {Adare}} \emph {et~al.} (\bibinfo {collaboration} {PHENIX Collaboration}),\
  }\bibfield  {title} {\bibinfo {title} {{Systematic study of azimuthal
  anisotropy in Cu+Cu and Au+Au collisions at ${\sqrt{s_{NN}}}=62.4$ and 200
  GeV}},\ }\href {https://doi.org/10.1103/PhysRevC.92.034913} {\bibfield
  {journal} {\bibinfo  {journal} {Phys. Rev. C}\ }\textbf {\bibinfo {volume}
  {92}},\ \bibinfo {pages} {034913} (\bibinfo {year} {2015})}\BibitemShut
  {NoStop}%
\bibitem [{\citenamefont {Adare}\ \emph {et~al.}(2016)\citenamefont {Adare}
  \emph {et~al.}}]{PPG183}%
  \BibitemOpen
  \bibfield  {author} {\bibinfo {author} {\bibfnamefont {A.}~\bibnamefont
  {Adare}} \emph {et~al.} (\bibinfo {collaboration} {PHENIX Collaboration}),\
  }\bibfield  {title} {\bibinfo {title} {{Measurements of directed, elliptic,
  and triangular flow in Cu+Au collisions at $\sqrt{{s}_{\mathit{NN}}}=200$
  GeV}},\ }\href {https://doi.org/10.1103/PhysRevC.94.054910} {\bibfield
  {journal} {\bibinfo  {journal} {Phys. Rev. C}\ }\textbf {\bibinfo {volume}
  {94}},\ \bibinfo {pages} {054910} (\bibinfo {year} {2016})}\BibitemShut
  {NoStop}%
\bibitem [{\citenamefont {J.}\ \emph {et~al.}(2023)\citenamefont {J.} \emph
  {et~al.}}]{PPGYura}%
  \BibitemOpen
  \bibfield  {author} {\bibinfo {author} {\bibfnamefont {A.~N.}\ \bibnamefont
  {J.}} \emph {et~al.} (\bibinfo {collaboration} {PHENIX Collaboration}),\
  }\bibfield  {title} {\bibinfo {title} {Measurement of
  $\ensuremath{\phi}$-meson production in {$\mathrm{Cu}+\mathrm{Au}$ collisions
  at $\sqrt{{s}_{NN}}=200$ GeV and $\mathrm{U}+\mathrm{U}$ collisions at
  $\sqrt{{s}_{NN}}=193$ GeV}},\ }\href
  {https://doi.org/10.1103/PhysRevC.107.014907} {\bibfield  {journal} {\bibinfo
   {journal} {Phys. Rev. C}\ }\textbf {\bibinfo {volume} {107}},\ \bibinfo
  {pages} {014907} (\bibinfo {year} {2023})}\BibitemShut {NoStop}%
\bibitem [{\citenamefont {Lacey}\ and\ \citenamefont
  {Taranenko}(2006)}]{Taranenko}%
  \BibitemOpen
  \bibfield  {author} {\bibinfo {author} {\bibfnamefont {R.~A.}\ \bibnamefont
  {Lacey}}\ and\ \bibinfo {author} {\bibfnamefont {A.}~\bibnamefont
  {Taranenko}},\ }\bibfield  {title} {\bibinfo {title} {{What do elliptic flow
  measurements tell us about the matter created in the little bang at RHIC?}},\
  }\href {https://doi.org/10.22323/1.030.0021} {\bibfield  {journal} {\bibinfo
  {journal} {PoS}\ }\textbf {\bibinfo {volume} {CFRNC2006}},\ \bibinfo {pages}
  {021} (\bibinfo {year} {2006})}\BibitemShut {NoStop}%
\bibitem [{\citenamefont {Wang}(2001)}]{Wang2001}%
  \BibitemOpen
  \bibfield  {author} {\bibinfo {author} {\bibfnamefont {X.-N.}\ \bibnamefont
  {Wang}},\ }\bibfield  {title} {\bibinfo {title} {Jet quenching and azimuthal
  anisotropy of large ${p}_{T}$ spectra in noncentral high-energy heavy-ion
  collisions},\ }\href {https://doi.org/10.1103/PhysRevC.63.054902} {\bibfield
  {journal} {\bibinfo  {journal} {Phys. Rev. C}\ }\textbf {\bibinfo {volume}
  {63}},\ \bibinfo {pages} {054902} (\bibinfo {year} {2001})}\BibitemShut
  {NoStop}%
\bibitem [{\citenamefont {Gyulassy}\ \emph {et~al.}(2004)\citenamefont
  {Gyulassy}, \citenamefont {Vitev}, \citenamefont {Wang},\ and\ \citenamefont
  {Zhang}}]{Gyulassy:2003mc}%
  \BibitemOpen
  \bibfield  {author} {\bibinfo {author} {\bibfnamefont {M.}~\bibnamefont
  {Gyulassy}}, \bibinfo {author} {\bibfnamefont {I.}~\bibnamefont {Vitev}},
  \bibinfo {author} {\bibfnamefont {X.-N.}\ \bibnamefont {Wang}},\ and\
  \bibinfo {author} {\bibfnamefont {B.-W.}\ \bibnamefont {Zhang}},\ }\bibfield
  {title} {\bibinfo {title} {{Jet quenching and radiative energy loss in dense
  nuclear matter}}\ }(\bibinfo {year} {2004})\ \bibinfo {note}
  {{CERN-2015-001}}\BibitemShut {NoStop}%
\bibitem [{\citenamefont {Adler}\ \emph {et~al.}(2007)\citenamefont {Adler}
  \emph {et~al.}}]{PRC_pi0_AUAU_2007}%
  \BibitemOpen
  \bibfield  {author} {\bibinfo {author} {\bibfnamefont {S.~S.}\ \bibnamefont
  {Adler}} \emph {et~al.} (\bibinfo {collaboration} {PHENIX Collaboration}),\
  }\bibfield  {title} {\bibinfo {title} {{A Detailed Study of High-$p_{T}$
  Neutral Pion Suppression and Azimuthal Anisotropy in Au+Au Collisions at
  $\sqrt{s_{NN}}=200$ GeV}},\ }\href
  {https://doi.org/10.1103/PhysRevC.76.034904} {\bibfield  {journal} {\bibinfo
  {journal} {Phys. Rev. C}\ }\textbf {\bibinfo {volume} {76}},\ \bibinfo
  {pages} {034904} (\bibinfo {year} {2007})}\BibitemShut {NoStop}%
\bibitem [{\citenamefont {Aidala}\ \emph {et~al.}(2018)\citenamefont {Aidala}
  \emph {et~al.}}]{ZHARKO}%
  \BibitemOpen
  \bibfield  {author} {\bibinfo {author} {\bibfnamefont {C.}~\bibnamefont
  {Aidala}} \emph {et~al.} (\bibinfo {collaboration} {PHENIX Collaboration}),\
  }\bibfield  {title} {\bibinfo {title} {{Production of $\pi^0$ and $\eta$
  mesons in Cu$+$Au collisions at $\sqrt{s_{_{NN}}}$=200 GeV}},\ }\href
  {https://doi.org/10.1103/PhysRevC.98.054903} {\bibfield  {journal} {\bibinfo
  {journal} {Phys. Rev. C}\ }\textbf {\bibinfo {volume} {98}},\ \bibinfo
  {pages} {054903} (\bibinfo {year} {2018})}\BibitemShut {NoStop}%
\bibitem [{\citenamefont {Acharya}\ \emph {et~al.}(2020)\citenamefont {Acharya}
  \emph {et~al.}}]{RADZEVICH}%
  \BibitemOpen
  \bibfield  {author} {\bibinfo {author} {\bibfnamefont {U.}~\bibnamefont
  {Acharya}} \emph {et~al.} (\bibinfo {collaboration} {PHENIX Collaboration}),\
  }\bibfield  {title} {\bibinfo {title} {{Production of $\pi^0$ and $\eta$
  mesons in U+U collisions at $\sqrt{s_{NN}}=$192 GeV}},\ }\href
  {https://doi.org/10.1103/PhysRevC.102.064905} {\bibfield  {journal} {\bibinfo
   {journal} {Phys. Rev. C}\ }\textbf {\bibinfo {volume} {102}},\ \bibinfo
  {pages} {064905} (\bibinfo {year} {2020})}\BibitemShut {NoStop}%
\bibitem [{\citenamefont {Afanasiev}\ \emph {et~al.}(2009)\citenamefont
  {Afanasiev} \emph {et~al.}}]{PRC_pi0_AUAU_2009}%
  \BibitemOpen
  \bibfield  {author} {\bibinfo {author} {\bibfnamefont {S.}~\bibnamefont
  {Afanasiev}} \emph {et~al.} (\bibinfo {collaboration} {PHENIX
  Collaboration}),\ }\bibfield  {title} {\bibinfo {title} {{High-$p_{T}$
  $\pi^{0}$ Production with Respect to the Reaction Plane in Au+Au Collisions
  at $\sqrt{s_{NN}}=200$ GeV}},\ }\href
  {https://doi.org/10.1103/PhysRevC.80.054907} {\bibfield  {journal} {\bibinfo
  {journal} {Phys. Rev. C}\ }\textbf {\bibinfo {volume} {80}},\ \bibinfo
  {pages} {054907} (\bibinfo {year} {2009})}\BibitemShut {NoStop}%
\bibitem [{\citenamefont {Adare}\ \emph {et~al.}(2010)\citenamefont {Adare}
  \emph {et~al.}}]{PRC_pi0_AUAU_2010}%
  \BibitemOpen
  \bibfield  {author} {\bibinfo {author} {\bibfnamefont {A.}~\bibnamefont
  {Adare}} \emph {et~al.} (\bibinfo {collaboration} {PHENIX Collaboration}),\
  }\bibfield  {title} {\bibinfo {title} {{Azimuthal anisotropy of neutral pion
  production in Au+Au collisions at $\sqrt{s_{NN}}=200$ GeV: Path-length
  dependence of jet quenching and the role of initial geometry}},\ }\href
  {https://doi.org/10.1103/PhysRevLett.105.142301} {\bibfield  {journal}
  {\bibinfo  {journal} {Phys. Rev. Lett.}\ }\textbf {\bibinfo {volume} {105}},\
  \bibinfo {pages} {142301} (\bibinfo {year} {2010})}\BibitemShut {NoStop}%
\bibitem [{\citenamefont {Adare}\ \emph
  {et~al.}(2013{\natexlab{a}})\citenamefont {Adare} \emph {et~al.}}]{PPG129}%
  \BibitemOpen
  \bibfield  {author} {\bibinfo {author} {\bibfnamefont {A.}~\bibnamefont
  {Adare}} \emph {et~al.} (\bibinfo {collaboration} {PHENIX Collaboration}),\
  }\bibfield  {title} {\bibinfo {title} {{Azimuthal anisotropy of $\pi^0$ and
  $\eta$ mesons in Au+Au collisions at $\sqrt{{s}_{NN}} =$ 200 GeV}},\ }\href
  {https://doi.org/10.1103/PhysRevC.88.064910} {\bibfield  {journal} {\bibinfo
  {journal} {Phys. Rev. C}\ }\textbf {\bibinfo {volume} {88}},\ \bibinfo
  {pages} {064910} (\bibinfo {year} {2013}{\natexlab{a}})}\BibitemShut
  {NoStop}%
\bibitem [{\citenamefont {Haque}\ \emph {et~al.}(2012)\citenamefont {Haque},
  \citenamefont {Lin},\ and\ \citenamefont {Mohanty}}]{U_orientation}%
  \BibitemOpen
  \bibfield  {author} {\bibinfo {author} {\bibfnamefont {M.~R.}\ \bibnamefont
  {Haque}}, \bibinfo {author} {\bibfnamefont {Z.-W.}\ \bibnamefont {Lin}},\
  and\ \bibinfo {author} {\bibfnamefont {B.}~\bibnamefont {Mohanty}},\
  }\bibfield  {title} {\bibinfo {title} {{Multiplicity, average transverse
  momentum, and azimuthal anisotropy in U+U collisions at $\sqrt{s_{NN}}=200$
  GeV using a multiphase transport model}},\ }\href
  {https://doi.org/10.1103/PhysRevC.85.034905} {\bibfield  {journal} {\bibinfo
  {journal} {Phys. Rev. C}\ }\textbf {\bibinfo {volume} {85}},\ \bibinfo
  {pages} {034905} (\bibinfo {year} {2012})}\BibitemShut {NoStop}%
\bibitem [{\citenamefont {Heinz}\ and\ \citenamefont
  {Kuhlman}(2005)}]{U_quenching}%
  \BibitemOpen
  \bibfield  {author} {\bibinfo {author} {\bibfnamefont {U.}~\bibnamefont
  {Heinz}}\ and\ \bibinfo {author} {\bibfnamefont {A.}~\bibnamefont
  {Kuhlman}},\ }\bibfield  {title} {\bibinfo {title} {Anisotropic flow and jet
  quenching in ultrarelativistic u+u collisions},\ }\href
  {https://doi.org/10.1103/PhysRevLett.94.132301} {\bibfield  {journal}
  {\bibinfo  {journal} {Phys. Rev. Lett.}\ }\textbf {\bibinfo {volume} {94}},\
  \bibinfo {pages} {132301} (\bibinfo {year} {2005})}\BibitemShut {NoStop}%
\bibitem [{\citenamefont {Allen}\ \emph {et~al.}(2003)\citenamefont {Allen}
  \emph {et~al.}}]{PHENIX_InnerDetectors}%
  \BibitemOpen
  \bibfield  {author} {\bibinfo {author} {\bibfnamefont {M.}~\bibnamefont
  {Allen}} \emph {et~al.} (\bibinfo {collaboration} {PHENIX Collaboration}),\
  }\bibfield  {title} {\bibinfo {title} {{PHENIX} inner detectors},\ }\href
  {https://doi.org/https://doi.org/10.1016/S0168-9002(02)01956-3} {\bibfield
  {journal} {\bibinfo  {journal} {Nucl. Instrum. Meth. A}\ }\textbf {\bibinfo
  {volume} {499}},\ \bibinfo {pages} {549} (\bibinfo {year}
  {2003})}\BibitemShut {NoStop}%
\bibitem [{\citenamefont {Poskanzer}\ and\ \citenamefont
  {Voloshin}(1998)}]{Methods_for_analyzing}%
  \BibitemOpen
  \bibfield  {author} {\bibinfo {author} {\bibfnamefont {A.~M.}\ \bibnamefont
  {Poskanzer}}\ and\ \bibinfo {author} {\bibfnamefont {S.~A.}\ \bibnamefont
  {Voloshin}},\ }\bibfield  {title} {\bibinfo {title} {Methods for analyzing
  anisotropic flow in relativistic nuclear collisions},\ }\href
  {https://doi.org/10.1103/PhysRevC.58.1671} {\bibfield  {journal} {\bibinfo
  {journal} {Phys. Rev. C}\ }\textbf {\bibinfo {volume} {58}},\ \bibinfo
  {pages} {1671} (\bibinfo {year} {1998})}\BibitemShut {NoStop}%
\bibitem [{\citenamefont {Barrette}\ \emph {et~al.}(1997)\citenamefont
  {Barrette} \emph {et~al.}}]{Flattening}%
  \BibitemOpen
  \bibfield  {author} {\bibinfo {author} {\bibfnamefont {J.}~\bibnamefont
  {Barrette}} \emph {et~al.} (\bibinfo {collaboration} {E877 Collaboration}),\
  }\bibfield  {title} {\bibinfo {title} {{Proton and pion production relative
  to the reaction plane in Au+Au collisions at 11A GeV/$c$}},\ }\href
  {https://doi.org/10.1103/PhysRevC.56.3254} {\bibfield  {journal} {\bibinfo
  {journal} {Phys. Rev. C}\ }\textbf {\bibinfo {volume} {56}},\ \bibinfo
  {pages} {3254} (\bibinfo {year} {1997})}\BibitemShut {NoStop}%
\bibitem [{\citenamefont {Adare}\ \emph {et~al.}(2012)\citenamefont {Adare}
  \emph {et~al.}}]{DirPhotAuAu}%
  \BibitemOpen
  \bibfield  {author} {\bibinfo {author} {\bibfnamefont {A.}~\bibnamefont
  {Adare}} \emph {et~al.} (\bibinfo {collaboration} {PHENIX Collaboration}),\
  }\bibfield  {title} {\bibinfo {title} {Observation of direct-photon
  collective flow in $\mathrm{Au}+\mathrm{Au}$ collisions at
  $\sqrt{{s}_{NN}}=200\text{ }\text{ }\mathrm{GeV}$},\ }\href
  {https://doi.org/10.1103/PhysRevLett.109.122302} {\bibfield  {journal}
  {\bibinfo  {journal} {Phys. Rev. Lett.}\ }\textbf {\bibinfo {volume} {109}},\
  \bibinfo {pages} {122302} (\bibinfo {year} {2012})}\BibitemShut {NoStop}%
\bibitem [{\citenamefont {Aidala}\ \emph {et~al.}(2014)\citenamefont {Aidala}
  \emph {et~al.}}]{FVTX}%
  \BibitemOpen
  \bibfield  {author} {\bibinfo {author} {\bibfnamefont {C.}~\bibnamefont
  {Aidala}} \emph {et~al.},\ }\bibfield  {title} {\bibinfo {title} {{The PHENIX
  Forward Silicon Vertex Detector}},\ }\href
  {https://doi.org/https://doi.org/10.1016/j.nima.2014.04.017} {\bibfield
  {journal} {\bibinfo  {journal} {Nucl. Instrum. Meth. A}\ }\textbf {\bibinfo
  {volume} {755}},\ \bibinfo {pages} {44} (\bibinfo {year} {2014})}\BibitemShut
  {NoStop}%
\bibitem [{\citenamefont {Adcox}\ \emph {et~al.}(2003)\citenamefont {Adcox}
  \emph {et~al.}}]{TrackingSystem}%
  \BibitemOpen
  \bibfield  {author} {\bibinfo {author} {\bibfnamefont {K.}~\bibnamefont
  {Adcox}} \emph {et~al.} (\bibinfo {collaboration} {PHENIX Collaboration}),\
  }\bibfield  {title} {\bibinfo {title} {{PHENIX central arm tracking
  detectors}},\ }\href@noop {} {\bibfield  {journal} {\bibinfo  {journal}
  {Nucl. Instrum. Meth. A}\ }\textbf {\bibinfo {volume} {499}},\ \bibinfo
  {pages} {489} (\bibinfo {year} {2003})}\BibitemShut {NoStop}%
\bibitem [{\citenamefont {Chiu}(2007)}]{MPC}%
  \BibitemOpen
  \bibfield  {author} {\bibinfo {author} {\bibfnamefont {M.}~\bibnamefont
  {Chiu}} (\bibinfo {collaboration} {PHENIX Collaboration}),\ }\bibfield
  {title} {\bibinfo {title} {{Single spin transverse asymmetries of neutral
  pions at forward rapidities in $\sqrt{s}=62.4$ GeV polarized proton
  collisions in PHENIX}},\ }\href {https://doi.org/10.1063/1.2750838}
  {\bibfield  {journal} {\bibinfo  {journal} {AIP Conf. Proc.}\ }\textbf
  {\bibinfo {volume} {915}},\ \bibinfo {pages} {539} (\bibinfo {year}
  {2007})}\BibitemShut {NoStop}%
\bibitem [{\citenamefont {Aphecetche}\ \emph {et~al.}(2003)\citenamefont
  {Aphecetche} \emph {et~al.}}]{EMCal}%
  \BibitemOpen
  \bibfield  {author} {\bibinfo {author} {\bibfnamefont {L.}~\bibnamefont
  {Aphecetche}} \emph {et~al.} (\bibinfo {collaboration} {PHENIX
  Collaboration}),\ }\bibfield  {title} {\bibinfo {title} {{PHENIX
  calorimeter}},\ }\href
  {https://doi.org/https://doi.org/10.1016/S0168-9002(02)01954-X} {\bibfield
  {journal} {\bibinfo  {journal} {Nucl. Instrum. Meth. A}\ }\textbf {\bibinfo
  {volume} {499}},\ \bibinfo {pages} {521} (\bibinfo {year}
  {2003})}\BibitemShut {NoStop}%
\bibitem [{\citenamefont {Adare}\ \emph
  {et~al.}(2013{\natexlab{b}})\citenamefont {Adare} \emph
  {et~al.}}]{PRC_AUAU_PRODUCTION_2013}%
  \BibitemOpen
  \bibfield  {author} {\bibinfo {author} {\bibfnamefont {A.}~\bibnamefont
  {Adare}} \emph {et~al.} (\bibinfo {collaboration} {PHENIX Collaboration}),\
  }\bibfield  {title} {\bibinfo {title} {{Neutral pion production with respect
  to centrality and reaction plane in Au+Au collisions at $\sqrt{s_{NN}}=200$
  GeV}},\ }\href {https://doi.org/10.1103/PhysRevC.87.034911} {\bibfield
  {journal} {\bibinfo  {journal} {Phys. Rev. C}\ }\textbf {\bibinfo {volume}
  {87}},\ \bibinfo {pages} {034911} (\bibinfo {year}
  {2013}{\natexlab{b}})}\BibitemShut {NoStop}%
\bibitem [{\citenamefont {Adare}\ \emph {et~al.}(2007)\citenamefont {Adare}
  \emph {et~al.}}]{pp}%
  \BibitemOpen
  \bibfield  {author} {\bibinfo {author} {\bibfnamefont {A.}~\bibnamefont
  {Adare}} \emph {et~al.} (\bibinfo {collaboration} {PHENIX Collaboration}),\
  }\bibfield  {title} {\bibinfo {title} {{Inclusive cross section and double
  helicity asymmetry for ${\ensuremath{\pi}}^{0}$ production in $p+p$
  collisions at $\sqrt{s}=200$ GeV: Implications for the polarized gluon
  distribution in the proton}},\ }\href
  {https://doi.org/10.1103/PhysRevD.76.051106} {\bibfield  {journal} {\bibinfo
  {journal} {Phys. Rev. D}\ }\textbf {\bibinfo {volume} {76}},\ \bibinfo
  {pages} {051106} (\bibinfo {year} {2007})}\BibitemShut {NoStop}%
\bibitem [{\citenamefont {Bass}\ \emph {et~al.}(2009)\citenamefont {Bass},
  \citenamefont {Gale}, \citenamefont {Majumder}, \citenamefont {Nonaka},
  \citenamefont {Qin}, \citenamefont {Renk},\ and\ \citenamefont
  {Ruppert}}]{rabJetQuench}%
  \BibitemOpen
  \bibfield  {author} {\bibinfo {author} {\bibfnamefont {S.~A.}\ \bibnamefont
  {Bass}}, \bibinfo {author} {\bibfnamefont {C.}~\bibnamefont {Gale}}, \bibinfo
  {author} {\bibfnamefont {A.}~\bibnamefont {Majumder}}, \bibinfo {author}
  {\bibfnamefont {C.}~\bibnamefont {Nonaka}}, \bibinfo {author} {\bibfnamefont
  {G.-Y.}\ \bibnamefont {Qin}}, \bibinfo {author} {\bibfnamefont
  {T.}~\bibnamefont {Renk}},\ and\ \bibinfo {author} {\bibfnamefont
  {J.}~\bibnamefont {Ruppert}},\ }\bibfield  {title} {\bibinfo {title}
  {Systematic comparison of jet energy-loss schemes in a realistic hydrodynamic
  medium},\ }\href {https://doi.org/10.1103/PhysRevC.79.024901} {\bibfield
  {journal} {\bibinfo  {journal} {Phys. Rev. C}\ }\textbf {\bibinfo {volume}
  {79}},\ \bibinfo {pages} {024901} (\bibinfo {year} {2009})}\BibitemShut
  {NoStop}%
\bibitem [{\citenamefont {Masui}\ \emph {et~al.}(2009)\citenamefont {Masui},
  \citenamefont {Mohanty},\ and\ \citenamefont {Xu}}]{Masui:2009qk}%
  \BibitemOpen
  \bibfield  {author} {\bibinfo {author} {\bibfnamefont {H.}~\bibnamefont
  {Masui}}, \bibinfo {author} {\bibfnamefont {B.}~\bibnamefont {Mohanty}},\
  and\ \bibinfo {author} {\bibfnamefont {N.}~\bibnamefont {Xu}},\ }\bibfield
  {title} {\bibinfo {title} {{Predictions of elliptic flow and nuclear
  modification factor from 200 GeV U+U collisions at RHIC}},\ }\href
  {https://doi.org/10.1016/j.physletb.2009.08.025} {\bibfield  {journal}
  {\bibinfo  {journal} {Phys. Lett. B}\ }\textbf {\bibinfo {volume} {679}},\
  \bibinfo {pages} {440} (\bibinfo {year} {2009})}\BibitemShut {NoStop}%
\bibitem [{hep()}]{hepdata}%
  \BibitemOpen
  \href@noop {} {}\bibinfo {howpublished}
  {https://www.hepdata.net/record/??????}\BibitemShut {Stop}%
\end{thebibliography}

%apsrev4-2.bst 2019-01-14 (MD) hand-edited version of apsrev4-1.bst
%Control: key (0)
%Control: author (8) initials jnrlst
%Control: editor formatted (1) identically to author
%Control: production of article title (0) allowed
%Control: page (0) single
%Control: year (1) truncated
%Control: production of eprint (0) enabled
%
 
\end{document}